\pdfoutput=1

\documentclass[11pt]{article}

\usepackage[final]{acl2025}

\usepackage{times}
\usepackage{latexsym}
\usepackage{algorithm}
\usepackage{algpseudocode}
\usepackage{xcolor}

\usepackage[T1]{fontenc}

\usepackage{microtype}

\usepackage{inconsolata}

\usepackage[utf8]{inputenc}
\usepackage{booktabs,multirow}
\usepackage{enumitem}
\usepackage{kotex} 
\usepackage{color}
\usepackage{rotating}
\usepackage{amsmath}
\usepackage{amsfonts}
\usepackage{pifont}
\usepackage{array}
\usepackage{amssymb}
\usepackage{algpseudocode}
\usepackage{algorithm}
\usepackage{float}
\usepackage{tcolorbox}
\usepackage{multicol}
\usepackage{lipsum}
\usepackage[normalem]{ulem}
\usepackage{soul}
\usepackage{listings}
\usepackage{amsmath} 
\usepackage{adjustbox}
\usepackage{tabularx}
\usepackage{booktabs}
\usepackage{multirow}
\usepackage{colortbl}
\usepackage{wrapfig}
\usepackage{placeins}
\tcbuselibrary{listings,breakable,skins}

\newcolumntype{Z}{>{\raggedright\arraybackslash}X}
\newcommand{\PromptText}[1]{\small\ttfamily #1}

\newtcblisting{graylisting}{
  colback=gray!10,
  colframe=gray!40,
  boxrule=0.4pt,
  arc=2pt,
  outer arc=2pt,
  listing only,
  breakable,
  enhanced,
  left=6pt,
  right=6pt,
  top=6pt,
  bottom=6pt,
  before skip=8pt,
  after skip=8pt,
  listing options={
    basicstyle=\ttfamily\small,
    breaklines=true,
    breakatwhitespace=false,
    columns=fullflexible,
    keepspaces=true
  }
}

\definecolor{customhighlight}{HTML}{DDB6E4}
\sethlcolor{customhighlight}

\colorlet{instructionbg}{gray!15}
\colorlet{questionbg}{gray!25}

\newcommand{\ie}{{\it i.e.}}

\newcommand{\ours}{GAREN}

\newcommand{\es}{\textcolor{orange}}
\title{Group-Aware Adaptive Retrieval for Evidence Navigation}

\author{June Park\thanks{\ \ Equal contribution.}, Jun Kwon\footnotemark[1], Jonghyo Kim, Jongwuk Lee\thanks{\ \ Corresponding author} \\
        Sungkyunkwan University, Republic of Korea\\  
        \texttt{\{pj00515, kwon04210, naye971012, jongwuklee\}@skku.edu}}

\begin{document}
\maketitle
\begin{abstract}

Reasoning-intensive retrieval addresses queries whose relevance cannot be identified by surface-level matching, thereby requiring multi-step reasoning. Because relevant documents rarely appear in the initial candidate set, retrieval systems suffer from the \emph{bounded recall problem}. Existing methods iteratively expand a candidate pool at the document level over a corpus graph, examining each neighbor in isolation and drifting toward a narrow region of the corpus. To address this problem, we propose \textit{\textbf{G}roup-Aware \textbf{A}daptive \textbf{R}etrieval for \textbf{E}vidence \textbf{N}avigation (\textbf{\ours{}})}, which explores the corpus graph through group-level expansion. 
\ours{} organizes documents into semantically coherent and distinguishable groups based on their connections in the corpus graph.
The information in each group indicates what can be accessed by expanding through it, providing guidance beyond individual document-level signals.
At each iteration, \ours\ uses a group-level navigator to identify promising expansion directions, retrieves documents from the selected groups, and applies a document-level reranker to evaluate the updated candidate set. Extensive experiments show that \ours\ achieves up to 8.0\% improvement over the strongest baseline on BRIGHT. 
The source code is available at \url{https://github.com/KJ12124/GAREN}

\end{abstract}

\section{Introduction}\label{sec:introduction}

Reasoning-intensive retrieval~\cite{BRIGHT} targets queries where relevance requires reasoning beyond lexical or semantic similarity. This indirect relevance makes relevant documents difficult to retrieve initially. Although stronger retrievers~\cite{ReasonIR} or query reformulation~\cite{ReDI} may improve recall, a fixed initial candidate set still struggles to capture documents whose relevance emerges only through multi-step reasoning. As a result, relevant documents missing from this initial candidate set are difficult to recover downstream, leaving the \textit{bounded recall problem} as a persistent challenge~\cite{SlideGAR}.

\begin{figure}[t]

\includegraphics[width=0.95\linewidth]{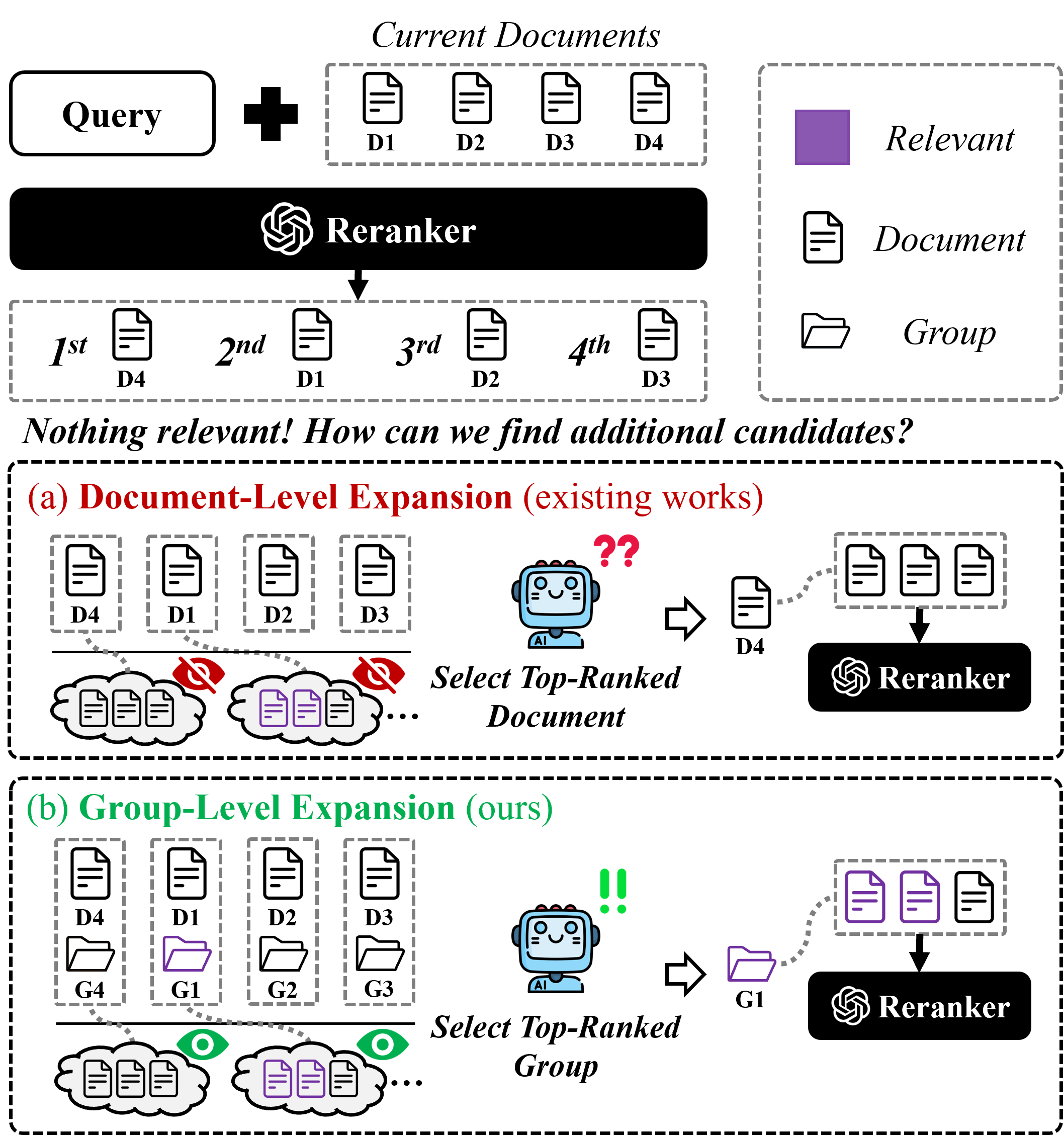}

\caption{Illustration of our motivation. While (a) existing methods expand based on individual documents, (b) \ours{} evaluates groups that reveal the documents accessible through each expansion direction.}\label{fig:fig_motivation}
\vskip -0.25in
\end{figure}

Recently, Adaptive Retrieval ~\cite{GAR} has emerged as a promising solution by adaptively expanding the candidate set. Existing methods~\cite{SlideGAR,RGS} leverage a corpus graph that connects similar documents, following the \textit{clustering hypothesis}~\cite{ClusteringHypothetis}, and expand the neighbors of documents assessed as relevant. At each step, the relevance of documents in the candidate set serves as evidence for deciding what to expand, enabling the expansion of additional relevant documents.

However, a document may not appear directly relevant to the query, even if expanding from it can lead to relevant documents~\cite{REPAIR}.
As shown in Figure~\ref{fig:fig_motivation}(a), when none of the current documents appear clearly relevant, document-level relevance signals provide little indication of which expansion direction is likely to lead toward relevant documents.
This limitation is especially critical in early retrieval stages, where the current candidate set provides only weak or unreliable evidence.
In such cases, early retrieval errors can propagate through the iterative expansion process, as documents retrieved along an incorrect direction introduce misleading signals for subsequent expansion decisions~\cite{REPAIR}.

This motivates our central research question:
\emph{How can expansion directions toward relevant documents be selected from weak and indirect signals?}
To address this question, we propose \textit{\textbf{G}roup-Aware \textbf{A}daptive \textbf{R}etrieval for \textbf{E}vidence \textbf{N}avigation} (\textbf{\ours{}}), a framework that evaluates expansion directions beyond individual document-level relevance.
For each expansion direction, it treats the documents that would be introduced by expanding in that direction as a \textit{group}.
As illustrated in Figure~\ref{fig:fig_motivation}(b), the group-level context indicates what additional information can be obtained by expanding in that direction, allowing the framework to assess whether the direction is worth exploring.
As a result, \ours{} can more reliably identify expansion directions that are likely to lead toward relevant documents, even when the current retrieval state provides only limited evidence.

Specifically, \ours{} consists of three steps. 
(i)~\textit{Group Construction} organizes densely connected documents into semantically coherent and distinguishable groups, enabling more informed comparison of potential expansion directions. 
It then summarizes the shared semantic context of each group into a representation that provides a coarse preview of the information reachable through that direction.
(ii) \textit{Group-Aware Adaptive Retrieval} then combines a document-level reranker and a group-level navigator: the reranker assesses the relevance of current candidate documents, while the navigator selects groups from which to expand next. 
Documents from the selected groups are added to the candidate set, which becomes the retrieval state for the next iteration. 
Within this process, we adopt an \textit{explore-then-exploit} strategy: the framework initially explores multiple groups to reduce uncertainty under limited evidence, and gradually focuses on more promising groups as reliable signals accumulate, reducing error propagation.
(iii) Following iterative retrieval, \textit{Group-Driven Evidence Propagation} refines candidate scores by allowing high-confidence documents to provide supporting evidence for other documents within the same group.

We summarize our key contributions as follows.
\begin{itemize}[leftmargin=*,topsep=0pt,parsep=0pt,partopsep=0pt,itemsep=0pt]

\item We introduce \textit{group} as a coarse-grained unit for expansion decisions in the corpus graph, allowing retrieval to assess where an expansion may lead.

\item We propose \ours{}, a group-aware adaptive retrieval framework that couples a coarse-grained navigator with a fine-grained reranker.

\item We show that \ours{} achieves strong performance on BRIGHT and remains effective on additional retrieval datasets.
\end{itemize}

\section{Related Work} \label{sec:related_work}

\subsection{Reasoning-Intensive Retrieval}\label{sec:Reasoning-Intensive Retrieval}

Recent reasoning-intensive retrieval benchmarks have highlighted scenarios in which relevant documents cannot be identified through lexical overlap or semantic similarity~\cite{BRIGHT}.
To address this challenge, prior work has incorporated reasoning into different stages of the retrieval pipeline.
One line of work enhances query interpretation using LLMs to decompose complex queries, generate reasoning chains, or rewrite queries with corpus feedback~\cite{ReDI,ThinkQE}.
Another line trains retrievers with reasoning-oriented supervision, encouraging retrieval models to capture indirect relevance~\cite{ReasonIR,RaDer}.
Others enrich documents offline with hypothetical information needs or query scenarios they can address, making implicit relevance explicit at indexing time~\cite{SPIKE,EnrichIndex}.
Complementary to these approaches, reranking methods perform reasoning over candidate documents to assess their relevance to the query~\cite{Reasonrank}.
These approaches are orthogonal to ours: query formulation, document enrichment, and stronger retrievers improve the initial retrieval state, while adaptive expansion recovers documents beyond the initial set.

\begin{figure*}[t]
\includegraphics[width=1.0\linewidth]{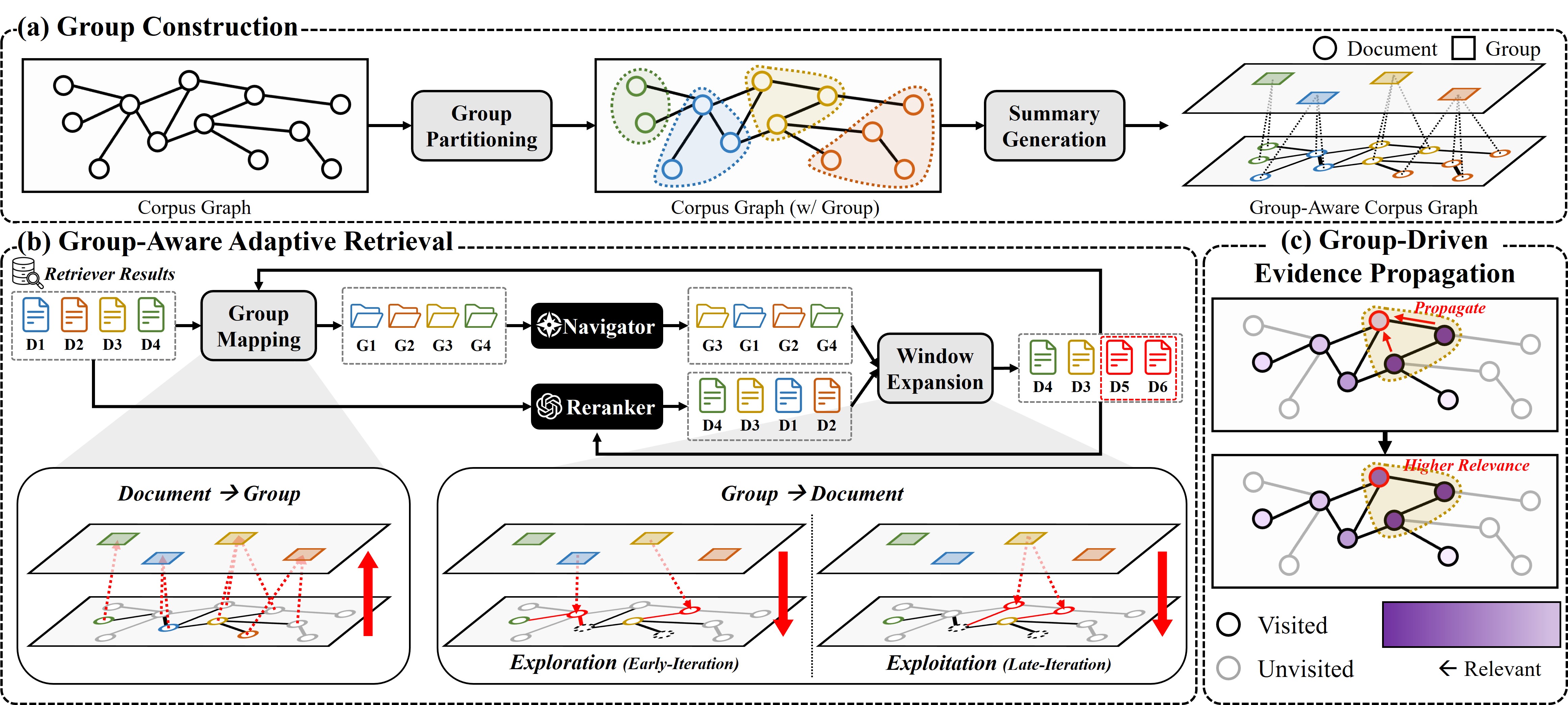}

\vskip -0.05in

\caption{
The overall framework of \ours{}. (a) The corpus graph is partitioned into semantic groups, each summarized by an LLM. (b) During retrieval, \ours{} iteratively reranks the current window and expands it with documents from groups identified by the navigator under an \textit{explore-then-exploit} strategy. (c) After retrieval, observed documents are re-scored using rank-weighted evidence from selected documents within the same group.
}
\label{fig:fig_main_framework}
\vskip -0.2in

\end{figure*}

\subsection{Adaptive Retrieval}\label{sec:Graph-based Retrieval}
Adaptive retrieval mitigates the \textit{bounded recall problem} by dynamically introducing additional candidates beyond the initial retrieval results.
Grounded in the clustering hypothesis, early graph-based methods construct document graphs and expand candidates through reranker-guided exploration~\cite{GAR,LADR}.
Subsequent methods further refine graph traversal by incorporating additional signals such as document-document relationships~\cite{Quam,ORE}.
More recent LLM-based methods use LLM-based scoring to guide graph traversal with richer semantic interpretation~\cite{SlideGAR,RGS}.
Other work incorporates intermediate reasoning steps as feedback for adaptive expansion or uses LLM preferences to construct corpus graphs~\cite{REPAIR,L2G}.
Despite these advances, relying primarily on signals from local candidate documents can make it difficult to assess broader expansion directions.

\subsection{Reasoning over Multiple Granularities}\label{sec:GRAPH_RAG} 

Prior work has shown that the choice of retrieval unit significantly affects both retrieval and downstream QA performance~\cite{DenseXRetrieval}.
Building on this insight, recent RAG methods organize corpora into hierarchical structures, enabling retrieval over both fine-grained evidence and higher-level abstractions~\cite{GraphRAG,RAPTOR,ArchRAG}.
In retrieval, recent work introduces an LLM-guided hierarchical retrieval framework that performs top-down traversal over a pre-constructed semantic tree ~\cite{LATTICE}.
In contrast, our approach is designed for iterative adaptive retrieval, where coarse-grained group signals and fine-grained document feedback guide expansion toward relevant documents more reliably.

\section{Preliminary}
\label{sec:preliminary}
\noindent\textbf{Listwise Reranking.}
Given a query $q$ and a candidate window $\mathcal{W}=(d_1,\dots,d_w)$ of size $w$, listwise reranking aims to reorder candidates so that documents relevant to $q$ are placed higher in the ranking.
We denote the listwise reranker by $\mathrm{Rerank}$, where
\begin{equation}
    \tilde{\mathcal{W}} = \mathrm{Rerank}(q, \mathcal{W}).
\end{equation}
The output $\tilde{\mathcal{W}}$ is an ordered list of the candidate documents.

\noindent\textbf{Adaptive Retrieval over Corpus Graphs.}
Reranking can only reorder documents that are already included in the candidate window.
To retrieve relevant documents outside this window, adaptive retrieval updates the window using a pre-constructed corpus graph $G=(D,E)$, where nodes are documents and edges capture semantic proximity.
Given a reranked window $\tilde{\mathcal{W}}$, adaptive retrieval preserves the top $h$ documents (the keep size) and appends $w-h$ graph-expanded candidates:
\begin{equation}
    \mathcal{X}
    =
    \mathrm{Expand}(q, \tilde{\mathcal{W}}, G, w-h),
\end{equation}
\begin{equation}
    \mathcal{W}_{\mathrm{next}}
    =
    \tilde{\mathcal{W}}[:h]
    \oplus
    \mathcal{X},
\end{equation}
where $\mathcal{X}$ denotes the graph-expanded candidates and $\oplus$ denotes list concatenation.
Existing methods typically realize $\mathrm{Expand}(\cdot)$ by deriving expansion anchors from the evidence in the current window. However, because they lack visibility into the information reachable through each potential expansion direction, they may fail to identify which direction leads toward relevant documents. 
We instead perform expansion at the group level, using group representations as coarse previews of the information reachable through alternative directions.

\renewcommand{\algorithmicrequire}{\textbf{Input:}}
\renewcommand{\algorithmicensure}{\textbf{Output:}}

\section{Proposed Method}
\label{sec:method}
We propose \ours{}, a novel group-aware adaptive retrieval framework, as illustrated in Figure~\ref{fig:fig_main_framework}.
\ours{} first partitions the corpus graph into semantically coherent groups (Section~\ref{sec:group_construction}).
During retrieval, \ours{} uses a navigator to select expansion directions at the group level (Section~\ref{sec:retrieval}).
After retrieval, \ours{} additionally applies a lightweight score refinement step that incorporates group evidence into the final ranking (Section~\ref{sec:boosting}).

\subsection{Group Construction}
\label{sec:group_construction}
Corpus graph expansion typically exposes only local document neighborhoods at each step, which makes it difficult to decide an expansion direction from a broader semantic context.
To provide a larger decision unit, we first construct semantically coherent document groups over the corpus graph.
This construction is performed offline on a pre-constructed corpus graph $G=(D,E)$, where nodes correspond to documents and edges capture embedding-based semantic proximity between them.

\noindent\textbf{Group Partitioning.}
For groups to serve as distinguishable expansion directions, each group should form a semantically coherent unit while remaining distinct from the others.
We realize this property through graph \textit{modularity}~\cite{Modularity}, which favors partitions with dense connections within groups and relatively sparse connections across groups.
Specifically, we apply Leiden community detection~\cite{Ledien} to optimize \textit{modularity} over the document-level semantic-proximity graph $G$, yielding a partition into densely connected document groups:

\begin{equation}
    \mathcal{P} = \{C_1,\dots,C_M\}.
\end{equation}
The groups in $\mathcal{P}$ form a partition of the corpus, and we denote the unique group containing document $d$ by $\Omega(d) \in \mathcal{P}$.
Detailed partitioning procedures and hyperparameters are described in Appendix~\ref{app:implementation_details}.

\noindent
\textbf{Summary Generation.}
For each group $C_m$, we use an LLM to generate a summary $S_m$ of its shared semantic context.
These summaries serve as coarse previews of information accessible through each expansion direction, without requiring all documents in a group to be examined individually.

\subsection{Group-Aware Adaptive Retrieval}
\label{sec:retrieval}

During retrieval, we use the constructed groups and summaries to guide expansion toward relevant documents. At each iteration, the navigator selects promising expansion directions from group-level information, and the reranker evaluates the documents exposed along those directions. Algorithm~\ref{algorithm:online} summarizes the overall procedure.

Since document-level reranking follows the retrieval formulation introduced in Section~\ref{sec:preliminary}, we focus below on the group-level expansion mechanism.
We first describe how the navigator scores candidate expansion directions, and then explain how our \textit{explore-then-exploit} strategy retrieves documents from these directions.
\begin{algorithm}[t]
\caption{Group-Aware Adaptive Retrieval}
\label{algorithm:online}
\small
\begin{algorithmic}[1]
\Require query $q$, initial ranked result $\mathcal{R}_0$, corpus graph $G$,
group map $\Omega$, neighborhood $N_k$, window size $w$, keep size $h$,
iterations $T$, switch iteration $\tau_{\mathrm{switch}}$
\Statex \quad $\triangleright$ \textcolor{blue}{\textbf{Window Initialization}}
\State $\mathcal{W}_1 \gets \mathcal{R}_0[:w]$
\For{$t = 1$ to $T$}
    \Statex \quad $\triangleright$ \textcolor{blue}{\textbf{Document Reranking}}
    \State $\tilde{\mathcal{W}}_t \gets \mathrm{Rerank}(q,\mathcal{W}_t)$
    \If{$t < T$}
        \Statex \quad $\triangleright$ \textcolor{blue}{\textbf{Group Mapping}}
        \State $\mathcal{G}^{\mathrm{cand}}_t \gets 
        \{\,\Omega(d') \mid d' \in \bigcup_{d \in \mathcal{W}_t} N_k(G,d)\,\}$
        \Statex \quad $\triangleright$ \textcolor{blue}{\textbf{Group Scoring}}
        \State $\mathcal{G}^{\mathrm{score}}_t \gets 
        \mathrm{Navigate}(q,\mathcal{G}^{\mathrm{cand}}_t)$
        \Statex \quad $\triangleright$ \textcolor{blue}{\textbf{Window Expansion}}
        \If{$t \leq \tau_{\mathrm{switch}}$}
            \State $\mathcal{X}_t \gets 
            \mathrm{Expand}^{\mathrm{explore}}(\mathcal{G}^{\mathrm{score}}_t,\, w-h)$
        \Else
            \State $\mathcal{X}_t \gets 
            \mathrm{Expand}^{\mathrm{exploit}}(\mathcal{G}^{\mathrm{score}}_t,\, w-h)$
        \EndIf
        \State $\mathcal{W}_{t+1} \gets 
        \tilde{\mathcal{W}}_t[:h] \oplus \mathcal{X}_t$
    \EndIf
\EndFor
\State \Return $\tilde{\mathcal{W}}_T$
\end{algorithmic}
\end{algorithm}

\subsubsection{Group-level Direction Scoring}
Let $\mathcal{W}_t$ be the document window at iteration $t$, and $N_k(G,d)$ the neighborhood of document $d$ in the pre-constructed $k$-NN corpus graph $G$.
Rather than scoring the entire group space at every iteration, we construct a set of candidate expansion directions around the current retrieval state:
\begin{equation}
\mathcal{G}^{\mathrm{cand}}_t =
\{\,\Omega(d') \mid d' \in \bigcup_{d \in \mathcal{W}_t} N_k(G,d)\,\}.
\end{equation}
Here, $\mathcal{W}_t$ specifies the current document-level retrieval state, while $\mathcal{G}^{\mathrm{cand}}_t$ comprises the group-level directions reachable from that state.
The navigator evaluates these candidate directions by examining the summaries of the groups in $\mathcal{G}^{\mathrm{cand}}_t$, which provide coarse previews of the information accessible through each direction:
\begin{equation}
\mathcal{G}^{\mathrm{score}}_t =
\mathrm{Navigate}(q,\mathcal{G}^{\mathrm{cand}}_t).
\end{equation}
The resulting $\mathcal{G}^{\mathrm{score}}_t$ provides scored group-level directions for the subsequent expansion step.
The navigator evaluates at most $K_g$ candidate groups at each iteration to keep the cost bounded.
The detailed scoring procedure is provided in Appendix~\ref{app:voting_caching}.

\subsubsection{Explore-then-Exploit Expansion}
Given the group scores in $\mathcal{G}^{\mathrm{score}}_t$, \ours{} selects groups for the next expansion.
In early iterations, the limited observed evidence makes it risky to concentrate expansion on a single high-scoring group, as an incorrect early
decision can misguide subsequent traversal.
We therefore adopt an explore-then-exploit strategy: \ours{} initially expands across multiple promising groups, and shifts to concentrating on high-scoring groups at iteration $\tau_{\mathrm{switch}}$, once relevant evidence has accumulated.

\begin{itemize}[leftmargin=*,topsep=0pt,parsep=0pt,partopsep=0pt,itemsep=0pt]
\item \textbf{Exploration}.
Early in the traversal, the limited number of observed documents makes the best expansion direction uncertain.
We therefore perform \textit{breadth-first expansion}, retrieving new documents from multiple high-scoring groups rather than concentrating on a single group.
Specifically, groups in $\mathcal{G}^{\mathrm{score}}_t$ are ranked by their navigator scores, and unobserved documents are selected from them in a round-robin manner.
This mitigates error propagation from incorrect early decisions and preserves opportunities to reach groups containing relevant evidence.
\item \textbf{Exploitation}.
As more documents are observed and reranked, the accumulated evidence provides a more reliable signal for identifying promising expansion directions.
We therefore perform \textit{depth-first expansion}, retrieving new documents preferentially from the highest-scoring groups.
Specifically, groups are visited in descending order of their navigator scores, and unobserved documents are selected from a higher-scoring group before moving to the next one.
This focuses subsequent traversal on groups supported by stronger accumulated evidence.
\end{itemize}

Finally, we obtain the expansion documents $\mathcal{X}_t$ following the detailed procedure in Appendix~\ref{app:voting_caching}, and construct the next retrieval window as:
\begin{equation}
    W_{t+1}
    =
    \tilde{W}_t[:h]
    \oplus
    \mathcal{X}_t.
\end{equation}

\subsection{Group-Driven Evidence Propagation}
\label{sec:boosting}
After the group-aware adaptive retrieval loop terminates, we apply a lightweight post-processing step called \emph{Group-Driven Evidence Propagation} over the observed document set.
This step is motivated by prior work that regularizes retrieval scores based on the clustering hypothesis~\citep{Rscore}.
Using the final ranking obtained after evidence has accumulated through iterative retrieval, we propagate rank-weighted support from selected documents to other observed documents in the same group.

Let $R_\mathrm{final}$ denote the final top-10 ranked results.
We first assign each selected document an RBP-style base weight~\citep{RBP}:
\begin{equation}
    b(d)
    =
    \begin{cases}
    (1-p)p^{\mathrm{rank}(d)-1}, & d \in R_\mathrm{final}, \\
    0, & \text{otherwise}.
    \end{cases}
\end{equation}
Here, $\mathrm{rank}(d)$ denotes the position of $d$ in
$R_\mathrm{final}$, and $p$ controls how quickly the weight decays with rank.
This rank-based decay assigns stronger evidence to higher-ranked documents,
allowing them to contribute more strongly to group-level support.

For each observed document, \emph{Group-Driven Evidence Propagation} combines its document-level evidence with the rank-weighted evidence accumulated in its
group:
\begin{equation}
    \mathrm{score}(d)
    =
    b(d)
    +
    \alpha
    \sum_{\substack{
        \Omega(d') = \Omega(d)
    }}
    b(d').
\end{equation}
Here, $\alpha$ controls how strongly evidence from neighboring selected documents is propagated.

Finally, we sort all observed documents by $\mathrm{score}(d)$ to obtain the final ranking.
This post-processing step helps promote potentially relevant documents belonging to evidence-rich semantic groups, but not sufficiently promoted by document-level reranking alone.

\begin{table*}[t]
\centering
\renewcommand{\arraystretch}{1.2}

\resizebox{\linewidth}{!}{

\begin{tabular}{l|ccccccc|cc|ccc|c}
\toprule
\multirow{2}{*}{\textbf{Method}}
& \multicolumn{7}{c|}{\textbf{StackExchange}}
& \multicolumn{2}{c|}{\textbf{Coding}}
& \multicolumn{3}{c|}{\textbf{Theorem-based}}
& \multirow{2}{*}{\textbf{Avg.}} \\

\cmidrule(lr){2-8}
\cmidrule(lr){9-10}
\cmidrule(lr){11-13}

& Bio. & Earth. & Econ. & Psy. & Rob. & Stack. & Sus.
& Leet. & Pony
& AoPS & TheoQ. & TheoT. & \\
\midrule

BM25
& 18.9 & 27.2 & 14.9 & 12.5 & 13.6 & 18.4 & 15.0
& 24.4 & 7.9
& 6.2 & 10.4 & 4.9
& 14.5 \\

\midrule
\multicolumn{14}{l}{\cellcolor{gray!15}\textit{Non-Reasoning-based Reranking}} \\
\midrule

Retrieve-and-Rerank
& 34.0 & 41.6 & 21.7 & 28.7 & 29.7 & 20.2 & 29.9
& \underline{17.5} & \underline{14.5}
& \underline{3.6}& 13.7 & 14.8
& 22.5 \\

SlideGAR
& 40.0 & \underline{42.1} & 25.9 & 32.0 & 28.2 & 19.9 & 32.7
& 14.4 & 12.4
& \textbf{4.1}& 11.6 & 22.6
& 23.8 \\

RGS
& \underline{43.3} & 42.0 & \underline{28.5} & \underline{35.4} & \underline{30.0} & \underline{22.9} & \textbf{34.5}
& \textbf{18.2}& 12.5
& 3.2 & \underline{15.0} & \textbf{32.0}
& \underline{26.4} \\

(Ours) \ours{}
& \textbf{45.3} & \textbf{43.6} & \textbf{30.3} & \textbf{38.5} & \textbf{31.6} & \textbf{25.2} & \underline{34.4}
& 16.7 & \textbf{20.4}
& \textbf{4.1}& \textbf{17.8} & \underline{31.0}
& \textbf{28.3} \\

\midrule
\multicolumn{14}{l}{\cellcolor{gray!15}\textit{Reasoning-based Reranking}} \\
\midrule

Retrieve-and-Rerank
& 34.7 & 42.5 & 23.8 & 29.0 & 31.6 & 23.2 & 32.0
& \textbf{23.4}& \underline{24.1}
& \underline{5.7} & 15.7 & 13.4
& 24.9 \\

SlideGAR
& 42.8 & 43.1 & 29.5 & 31.3 & 31.1 & 25.0 & \underline{36.5}
& 17.3 & 20.5
& 3.6 & 14.3 & 16.8
& 26.0 \\

RGS
& \underline{46.6} & \underline{46.0} & \underline{32.4} & \underline{35.3} & \underline{33.3} & \underline{29.4} & 35.9
& \underline{21.7} & 19.9
& 3.3 & \underline{17.8} & \underline{25.7}
& \underline{28.9} \\

REPAIR
& 44.6 & 43.7 & 32.3 & 33.7 & 32.0 & 29.2 & 36.1
& 20.3 & \underline{24.3}
& 3.6 & 15.3 & 25.3
& 28.4 \\

(Ours) \ours{}
& \textbf{47.1} & \textbf{46.5} & \textbf{34.1} & \textbf{40.6} & \textbf{37.1} & \textbf{31.0} & \textbf{37.8}
& 21.1 & \textbf{25.0}
& \textbf{6.2} & \textbf{18.6} & \textbf{29.9}
& \textbf{31.2} \\

\bottomrule
\end{tabular}
}
\caption{
nDCG@10 performance of various methods and reranking settings on BRIGHT.
The best performance is marked in \textbf{bold}, and the second-best is \underline{underlined} within the same reranking prompt setting.
Across all BRIGHT queries, GAREN significantly outperforms the best-performing baseline in both reranking settings under a query-level paired t-test ($p < 0.05$).
}
\label{tab:main_results}
\vskip -0.1in
\end{table*}

\section{Experimental Setup}
\label{sec:exp_setup}

\noindent
\textbf{Datasets}. 
We evaluate \ours{} on three retrieval benchmarks: BRIGHT~\cite{BRIGHT}, R2MED~\cite{R2MED}, and BEIR~\cite{BEIR}.
BRIGHT is a reasoning-intensive benchmark across diverse domains, and R2MED evaluates reasoning-driven medical retrieval.
BEIR provides diverse retrieval tasks; 
following prior work~\cite{RankGPT}, we evaluate on eight BEIR datasets.
For metrics, we report nDCG@10 and Recall@100 (R@100), where R@100 measures recall over the documents observed by the reranker.

\noindent
\textbf{Baselines}. 
We compare our method with four baselines.
(i) \textbf{Retrieve-and-Rerank} performs sliding-window reranking~\cite{RankGPT} within a fixed first-stage retrieval pool, without expanding the pool through the corpus graph.
(ii) \textbf{SlideGAR} ~\cite{SlideGAR} alternates between the retrieval pool and the corpus graph by expanding the window with document-level graph neighbors.
(iii) \textbf{RGS} ~\cite{RGS} performs adaptive retrieval on an ANN corpus graph.
(iv) \textbf{REPAIR} ~\cite{REPAIR} performs graph-based adaptive retrieval guided by intermediate reasoning steps.

\noindent
\textbf{Implementation Details}.
For all methods, we use BM25~\cite{SIGIR/RobertsonW94/BM25} for first-stage retrieval, 
construct the corpus graph using Qwen3-Embedding-4B~\cite{qwen3embedding}, and use 
Qwen3-Next-80B-A3B-Instruct-FP8~\cite{qwen3technicalreport} as the reranker with a window size 
of $w=20$, a keep size of $h=10$, and a total reranking budget of 100 documents.
All experiments are conducted with the vLLM~\cite{vllm} inference engine, and we report results averaged over 5 runs due to the variance of vLLM inference. 
Unless otherwise specified, we use the non-reasoning prompt setting. For \ours{}, we additionally use Qwen3-Next-80B-A3B-Instruct-FP8 for offline group summary generation and Qwen3-Reranker-4B as the navigator, with a group scoring budget of $K_g=20$ and $\tau_{\mathrm{switch}}=4$. Additional hyperparameters are in Appendix~\ref{app:implementation_details}.
\section{Results and Analysis}
\label{sec:results}

\begin{table}[t]
\centering
\renewcommand{\arraystretch}{1.2}
\resizebox{\columnwidth}{!}{
\begin{tabular}{lcccc}
\toprule
\multirow{2}{*}{Method} & \multicolumn{2}{c}{R2MED} & \multicolumn{2}{c}{BEIR} \\
\cmidrule(lr){2-3} \cmidrule(lr){4-5}
 & nDCG@10 & R@100 & nDCG@10 & R@100 \\
\midrule

BM25     
& 15.1 & 44.6 & 43.4 & 43.7 \\

Retrieve-and-Rerank         
& 35.0 & 44.6 & 51.0 & 43.7 \\

SlideGAR                    
& 42.7 & 62.9 & 51.3 & \textbf{48.6} \\

RGS                         
& \underline{44.5} & \underline{63.7} & \underline{52.0} & 43.7 \\

\midrule

(Ours) \ours{}              
& \textbf{45.1} & \textbf{67.6} & \textbf{52.4} & \underline{47.4} \\

\bottomrule
\end{tabular}
}

\caption{
Performance comparison on R2MED and BEIR.
The best performance is marked in \textbf{bold}, and the second-best is \underline{underlined}.
}
\label{tab:BEIR_R2MED}
\vskip -0.2in
\end{table}

\subsection{Main Results}
\noindent\textbf{Effectiveness on Reasoning-Intensive Retrieval.}
Table~\ref{tab:main_results} compares \ours{} with baselines on BRIGHT under both non-reasoning and reasoning prompting settings.
(i) In the non-reasoning setting, \ours{} improves average nDCG@10 over Retrieve-and-Rerank by 25.8\%, demonstrating the benefit of expanding beyond the initial retrieval pool, where relevant documents may be missing.
\ours{} also outperforms adaptive retrieval baselines such as SlideGAR and RGS by 18.9\% and 7.2\%, indicating that group-level expansion provides more effective expansion directions than document-level traversal based on local implicit signals.
(ii) In the reasoning setting, \ours{} again achieves the strongest overall performance, improving over Retrieve-and-Rerank by 25.3\%. Although REPAIR improves expansion decisions using reasoning-based anchors, \ours{} achieves stronger performance by explicitly comparing group-level expansion directions.
(iii) Across both settings, reranking-based methods show limited effectiveness on AoPS and LeetCode, and adaptive expansion provides little additional gain. In these subsets, the reranker and the embedding model often fail to identify the underlying algorithmic or mathematical techniques that connect relevant documents, favoring surface-level overlap instead. The effectiveness of our method is thus expected to improve as stronger rerankers and embedding models become available.

\noindent\textbf{Evaluation on R2MED and BEIR.}
As shown in Table~\ref{tab:BEIR_R2MED}, \ours{} achieves competitive performance on both benchmarks.
These results indicate that group-level guidance generalizes beyond BRIGHT to both medical reasoning and traditional retrieval settings.
The improvement margins are relatively modest compared with BRIGHT, because first-stage retrieval can already recover gold documents on these benchmarks.

\subsection{Document-Level vs.\ Group-Level Expansion}
\label{sec:DOC_GROUP}

\begin{figure}[t]
\includegraphics[width=1.0\linewidth]{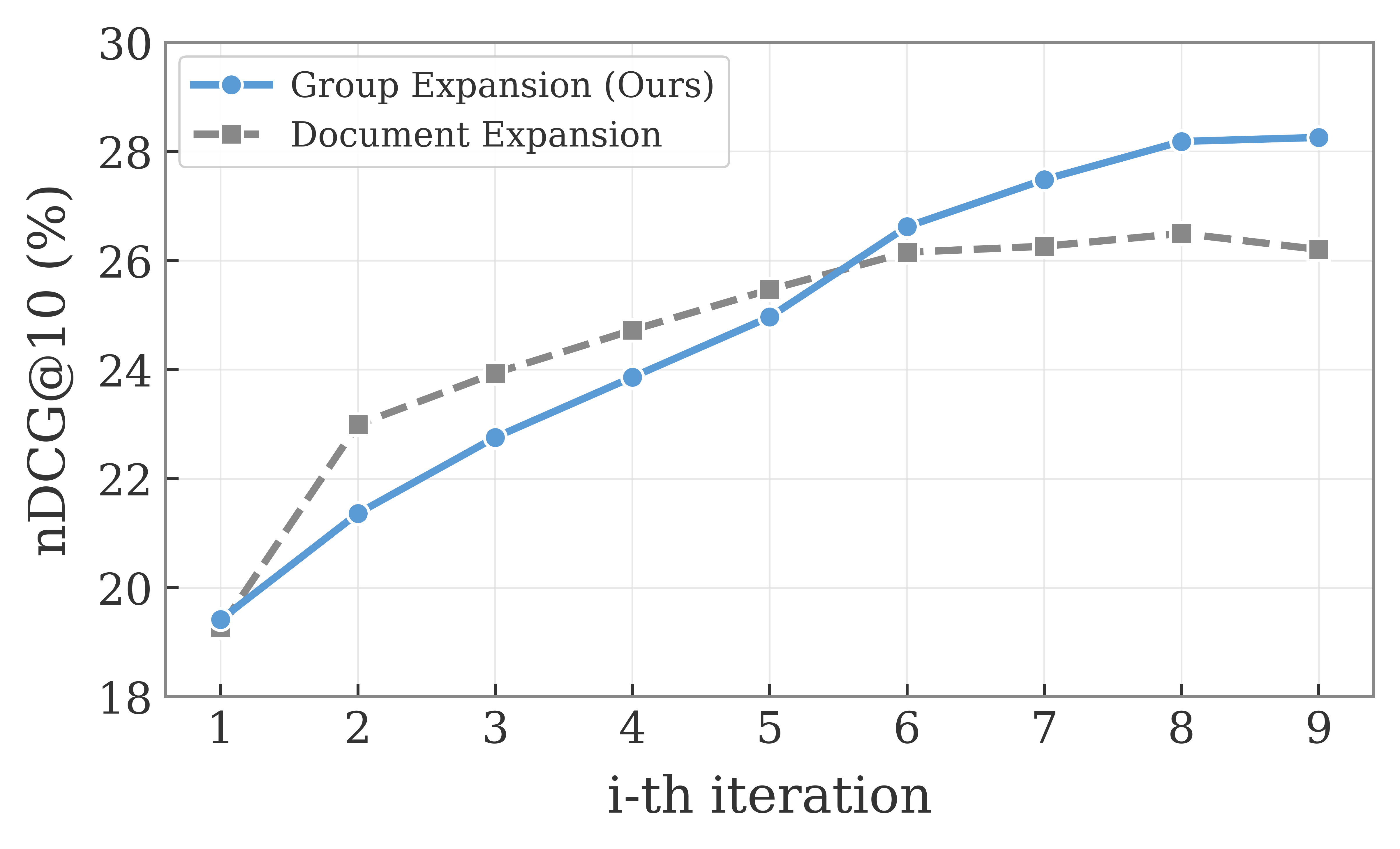}
\caption{Performance comparison of Document Expansion and Group Expansion. Unlike Document Expansion, Group Expansion consistently improves performance as the number of iterations increases.}\label{fig:DocVsGroup}
\vskip -0.2in
\end{figure}

We examine the benefit of group-level over document-level expansion by comparing \ours{} against a document-level variant that directly scores neighboring documents $N_k(d)$
connected to the current window with the navigator.~\footnote{For each step, the document-level variant can observe up to $\text{window size} \times \#\text{edges} = 20 \times 16 = 320$ documents.}
As shown in Figure~\ref{fig:DocVsGroup}, the document-level variant improves rapidly in early iterations but quickly saturates.
Direct neighbor scoring is effective near the current window, but has limited ability to discover relevant documents beyond immediate graph neighbors.
In contrast, group-level expansion improves more steadily and eventually surpasses the document-level variant.
By evaluating semantic groups, \ours{} selects expansion directions using broader contextual signals beyond immediate neighbors.
We further analyze this gap in Appendix~\ref{sec:DOC_GROUP_ANALYSIS}, showing that the advantage of group-level expansion grows as gold documents become more distant, which is structurally explained by the group graph substantially reducing the graph distance to gold documents.
\subsection{Gold Document Location Analysis}
\label{sec:gold_location}
\begin{figure}[t]
\includegraphics[width=1.0\linewidth]{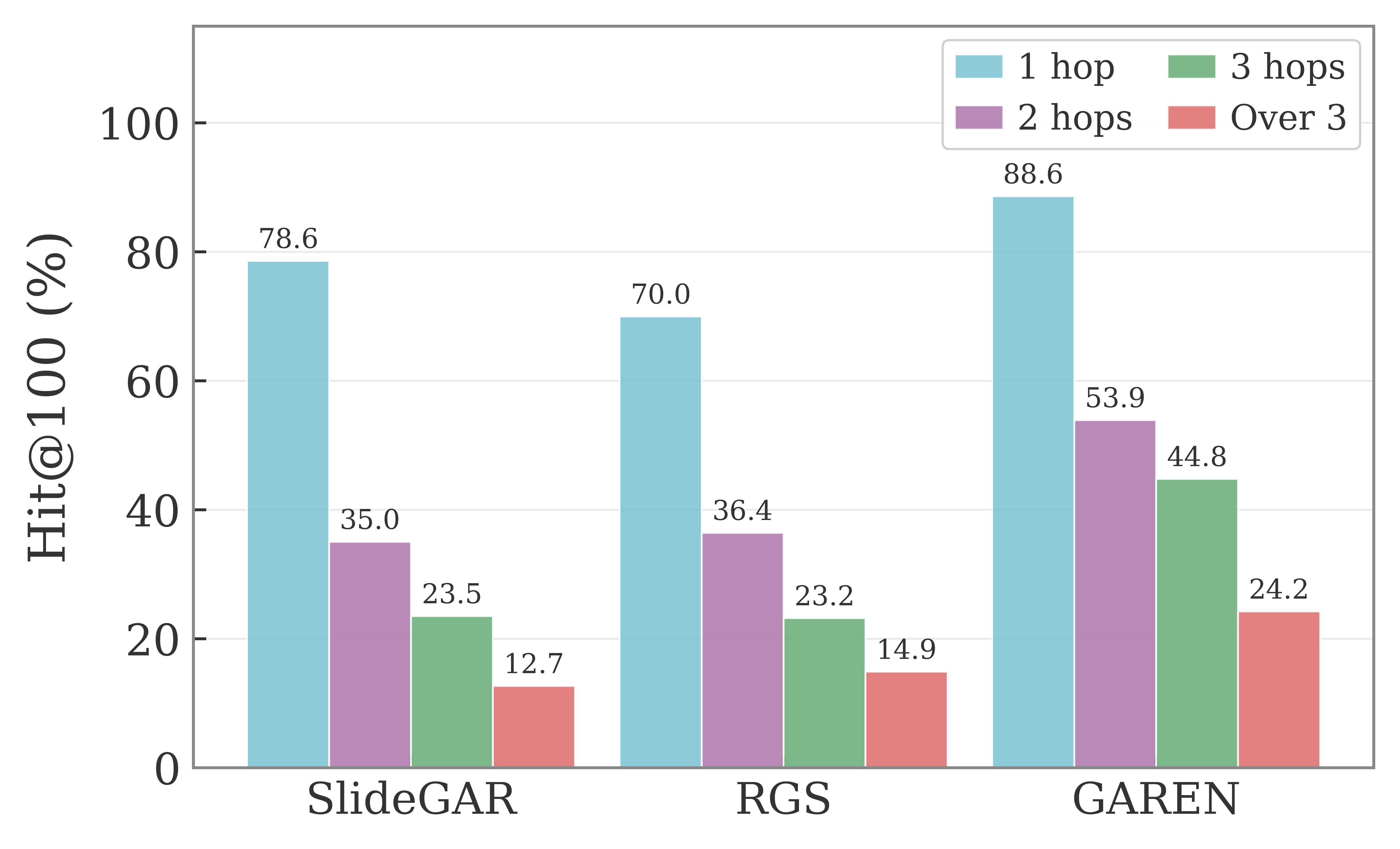}

\caption{
Hit@100 comparison between \ours{} and baseline methods by graph-hop distance to the nearest gold document on BRIGHT.
The gap between GAREN and the baselines widens as gold documents become farther from the initial retrieval results.
}\label{fig:reachability_by_distance}\label{fig:GoldDocLocaion_main}
\vskip -0.1in
\end{figure}

\begin{table}[t]

\centering
\renewcommand{\arraystretch}{1.2}
\resizebox{\columnwidth}{!}{
\begin{tabular}{ccc|cc}
\toprule
\multirow{2}{*}{Strategy} 
& \multirow{2}{*}{Propagation} 
& \multirow{2}{*}{Navigator} 
& \multicolumn{2}{c}{BRIGHT Avg.} \\
\cmidrule(lr){4-5}
& & & nDCG@10 & R@100 \\
\midrule
\textit{Explore-then-Exploit}     & \ding{51} & \ding{51} & \textbf{28.3} & \textbf{53.1} \\
\midrule
\textit{Exploration}  & \ding{51} & \ding{51} & 27.0 & 50.2 \\
\textit{Exploitation} & \ding{51} & \ding{51} & 27.8 & 51.1 \\
\textit{Explore-then-Exploit}     & \ding{55}    & \ding{51} & 27.9 & \textbf{53.1} \\
\textit{Explore-then-Exploit}     & \ding{51}    & \ding{55} & 21.2 & 38.4 \\
\bottomrule
\end{tabular}
}
\caption{Ablation study of \ours{} on BRIGHT. We analyze the effects of navigator, expansion strategy, and propagation.}
\label{tab:ablation}
\vskip -0.2in
\end{table}

We analyze whether \ours{} retrieves gold documents far from the initial documents.
On BRIGHT, we group queries by the minimum graph-hop distance from the BM25 top-20 set to a gold document, and report Hit@100, indicating whether at least one gold document is observed during retrieval. We focus on queries whose gold documents lie outside the BM25 top-20 set and must therefore be reached through expansion.
As shown in Figure~\ref{fig:GoldDocLocaion_main}, \ours{} observes more gold documents than graph-based baselines across all hop distances.
SlideGAR performs well only for nearby gold documents, while RGS reaches farther regions through ANN graph traversal but shows a trade-off between nearby and distant cases.
\ours{} maintains strong performance across distances, supporting our hypothesis that group-level expansion can reach gold documents beyond local document-level traversal.

\begin{figure}[t]
\includegraphics[width=1.0\linewidth]{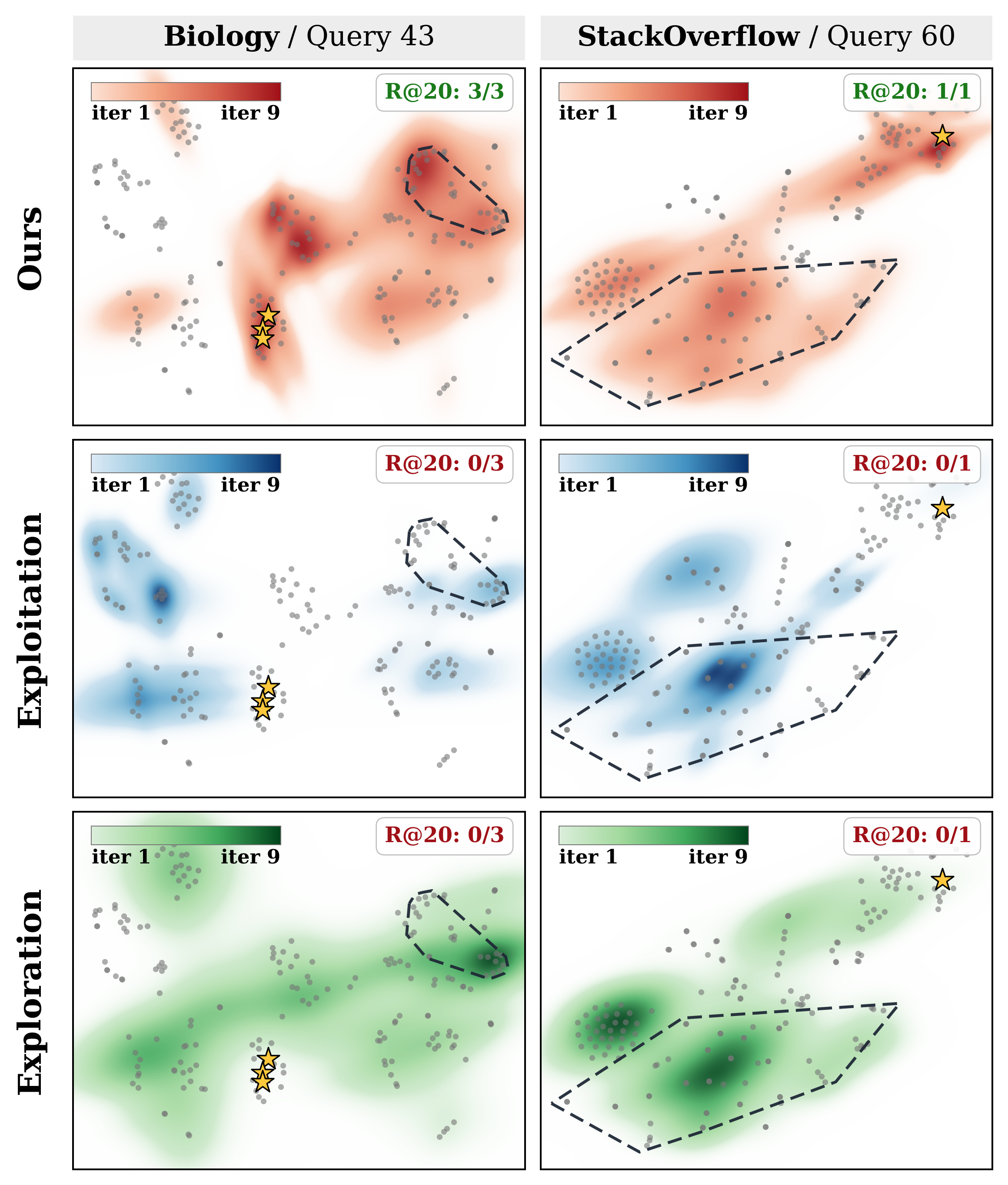}
\caption{
Case study on expansion strategies. \ours{} successfully retrieves the gold documents through an explore-then-exploit strategy. The dashed area denotes the initial region at iteration 1.}
\label{fig:CaseStudy_explore_exploitcation}
\vskip -0.1in
\end{figure}

\subsection{Ablation Study}
We construct several variants to assess the impact of each component in \ours{} and evaluate them on BRIGHT, as shown in Table~\ref{tab:ablation}.

\noindent\textbf{Explore-then-Exploit Strategy.}
Using a single expansion strategy degrades performance: \textit{Exploration} fails to concentrate on promising directions in later iterations, while \textit{Exploitation} commits too early before sufficient evidence emerges to identify directions leading to relevant documents.
These results confirm that our \textit{explore-then-exploit} strategy is effective for reaching relevant documents throughout the retrieval process.

\noindent\textbf{Group-Driven Evidence Propagation.}
Removing this component degrades the final ranking quality, showing that sharing evidence from high-confidence documents with other members of the same group helps prioritize relevant documents.

\noindent\textbf{Navigator Guidance.}
We replace navigator-based group scoring with a connectivity-based heuristic while maintaining all other components. 
Specifically, candidate groups are scored based on their connectivity to the neighboring documents in the current window.
This replacement reduces nDCG@10 and R@100 by 25.1\% and 27.7\%, respectively, highlighting the importance of semantic group evaluation for expansion.

\begin{figure}[t]
\includegraphics[width=1.0\linewidth]{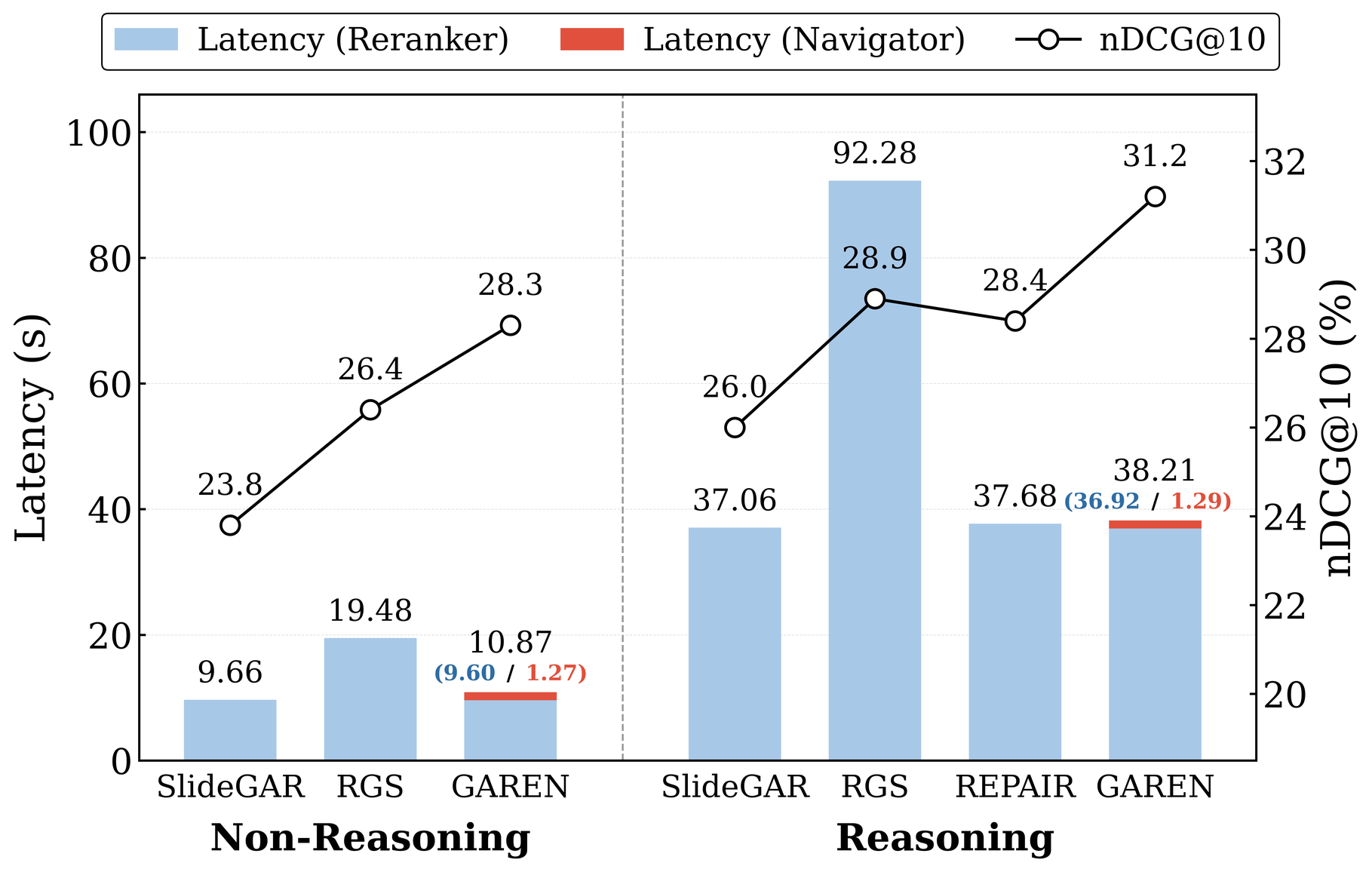}
\caption{
Retrieval effectiveness and latency comparison on BRIGHT, measured on 30 queries sampled from 
each subset. \ours{} achieves strong retrieval performance with competitive latency under both 
prompting settings.
Values in parentheses indicate the latency breakdown of the reranker and navigator, respectively.
}\label{fig:latency_accuracy_curve}
\vskip -0.1in
\end{figure}

\subsection{Case Study for Expansion Strategy}
We conduct a case study to visualize how different expansion strategies traverse the corpus, as shown in Figure~\ref{fig:CaseStudy_explore_exploitcation}.
\emph{Exploitation} focuses only on the highest-scoring groups, remaining near the seed set and failing to reach the gold documents.
\emph{Exploration} spreads expansion across many groups, exploring diverse directions and approaching the gold documents but failing to concentrate enough to retrieve them.
In contrast, \ours{} first explores diverse directions broadly, and then concentrates subsequent expansion on the most promising direction, successfully retrieving the gold documents.
This confirms that the \textit{explore-then-exploit} strategy is necessary to both \emph{explore} promising directions and \emph{exploit} them to retrieve gold documents.

\subsection{Efficiency Analysis}\label{sec:efficiency}
To evaluate the efficiency of \ours{}, we measure latency on BRIGHT. We sample 30 queries from each subset, resulting in 360 queries in total. 
The reported nDCG scores are computed over the full set of queries.
Experiments are conducted on two A100 GPUs. All queries are processed sequentially, and we report the average latency.
As shown in Figure~\ref{fig:latency_accuracy_curve}, \ours{} achieves high accuracy at competitive latency under both prompting settings. 
These results indicate that the navigator, being substantially smaller than the reranker, introduces only limited overhead while enabling group-level navigation. 
This design choice is further supported in Appendix~\ref{sec:Pipeline_Robustness}, where scaling the reranker yields substantially larger gains than scaling the navigator, indicating that a lightweight navigator is sufficient for group-level direction assessment.
Moreover, since the reranker and navigator operate independently on the current window, they can be executed in parallel in practical deployments to further reduce latency.

\section{Conclusion}\label{sec:conclusion}
In this paper, we propose \textbf{\ours{}}, a group-aware adaptive retrieval framework for addressing the  \textit{bounded recall problem} in reasoning-intensive retrieval.
Existing adaptive retrieval methods rely on local signals from the current candidates to determine expansion directions, which can be ineffective when the initial retrieval state provides limited or unreliable clues.
To mitigate this, \ours{} organizes the corpus graph into coarse-grained \textit{groups} and coordinates a group-level navigator with a document-level reranker through an \textit{explore-then-exploit} strategy.
A \textit{group-driven evidence propagation} step further consolidates evidence around high-confidence candidates.
Experiments demonstrate that \ours{} outperforms strong adaptive retrieval baselines, particularly when relevant documents are distant from the initial retrieval pool.
\section{Limitations}\label{sec:limitation}

We discuss the limitations of \ours{} as follows:

\noindent\textbf{Group Construction Strategy.}
We construct document groups using community-detection-based partitioning over the corpus graph, as our primary focus is on leveraging group-level structure for adaptive retrieval.
While this design performs effectively in our experiments, more advanced construction strategies may produce groups that provide more informative expansion directions.
For example, learned partitioning, alternative community detection algorithms, or corpus graphs built from stronger embeddings could further improve retrieval performance.

\noindent\textbf{Fixed Explore-then-Exploit Transition.}
We transition from exploration to exploitation using a fixed parameter $\tau_{\mathrm{switch}}$.
This schedule performs effectively across our experiments by encouraging broad exploration in early iterations and more focused expansion in later iterations.
Incorporating query-specific characteristics or intermediate retrieval signals could enable a more adaptive transition that better identifies when sufficient evidence has accumulated for focused expansion.

\section*{Ethics Statement}

This work adheres to the ACL's ethical guidelines. All scientific resources were obtained under permissive licenses and used for their intended research purposes.

\section*{Acknowledgments}
This work was supported by the Institute of Information \& communications Technology Planning \& Evaluation (IITP) grant and the National Research Foundation of Korea (NRF) grant funded by the Korea government (MSIT) (No. NRF-RS-2025-00564083, IITP-RS-2022-II220680, IITP-2025-RS-2020-II201821, IITP-2026-RS-2024-00437633) and the SEMES-SKKU collaboration funded by SEMES, each contributing 20\% to this research.

\bibliography{references}

\appendix

\newpage

\section{Additional Setup}
\label{app:implementation_details}

\paragraph{Corpus Graph.}
We construct the corpus graph by connecting each document to its top 16 most similar documents based on Qwen3-Embedding-4B representations. For RGS, we follow the original implementation and use DiskANN~\cite{DiskANN} for ANN graph construction.

\paragraph{Group Construction.}
We partition the corpus graph using the Leiden community detection algorithm.
We set the resolution parameter to $\gamma=1.0$ and the randomness parameter to $\theta=0.001$.
To encourage fine-grained groups, we use a re-partitioning threshold of $\bar{s}=20$: groups exceeding this threshold are recursively re-partitioned whenever Leiden identifies a non-trivial subdivision.
Consequently, larger groups may remain when no further meaningful partition is identified.
After partitioning, each group is summarized once offline using the summary prompt shown in Appendix~\ref{app:prompts}.

\paragraph{Group-Driven Evidence Propagation.}
We set the RBP decay parameter to $p=0.6$ and the neighbor propagation weight to $\alpha=0.01$ in the non-reasoning reranking setting, and $p=0.5$ and $\alpha=0.01$ in the reasoning setting.
These values were selected via greedy search on the development splits of NFCorpus (324 queries) and FiQA (500 queries) in BEIR.

\section{Extended Experiments}
\subsection{Pipeline Robustness}\label{sec:Pipeline_Robustness}
We examine whether \ours{} remains effective across various configurations of the retrieval pipeline by varying the first-stage retrieval, the corpus graph embedding, and the reranker and navigator models.

\noindent\textbf{First-Stage Retrieval.}
We evaluate whether \ours{} remains effective when the initial retrieval pool is strengthened.
To this end, we replace the BM25 first-stage retriever with ReasonIR~\cite{ReasonIR}, a retriever trained for reasoning-intensive retrieval, and further strengthen the initial pool with GPT-4 queries from BRIGHT~\cite{BRIGHT}. All methods are evaluated under the reasoning prompt setting.
As shown in Table~\ref{tab:appendix_reasonir_ndcg}, since the initial pool already contains many relevant documents, fewer documents remain to be recovered through expansion and the margins between methods are narrower than in the BM25 setting. Nevertheless, \ours{} still achieves the best performance under the stronger pool, showing that group-level guidance 
reliably identifies useful expansion directions even when the room for expansion is limited.
\begin{table*}[t]
\centering
\renewcommand{\arraystretch}{1.2}

\resizebox{\linewidth}{!}{

\begin{tabular}{l|ccccccc|cc|ccc|c}
\toprule
\multicolumn{14}{l}{\textit{\textbf{First-stage retrieval:} BM25 $\rightarrow$ ReasonIR with GPT-4 queries}} \\
\midrule

\multirow{2}{*}{\textbf{Method}}
& \multicolumn{7}{c|}{\textbf{StackExchange}}
& \multicolumn{2}{c|}{\textbf{Coding}}
& \multicolumn{3}{c|}{\textbf{Theorem-based}}
& \multirow{2}{*}{\textbf{Avg.}} \\

\cmidrule(lr){2-8}
\cmidrule(lr){9-10}
\cmidrule(lr){11-13}

& Bio. & Earth. & Econ. & Psy. & Rob. & Stack. & Sus.
& Leet. & Pony
& AoPS & TheoQ. & TheoT. & \\ 
\midrule
ReasonIR
& 43.0 & 43.2 & 33.1 & 39.7 & 20.9 & 30.4 & 27.4
& 31.5& 19.7 & 7.4& 34.0& 37.0& 30.6 \\

\midrule
\multicolumn{14}{l}{\cellcolor{gray!15}\textit{Reasoning-based Reranking}} \\
\midrule

Retrieve-and-Rerank
& 51.5 & 45.7 & 34.2 & 43.0 & 30.6 & 30.5 & 38.9
& 18.6 & \textbf{29.5}
& 4.9 & 30.2 & 36.2
& 32.8 \\

SlideGAR
& 53.2 & 46.3 & 34.0 & 43.8 & 30.0 & 29.4 & 37.9
& 17.2 & \underline{26.7}
& 4.3 & 29.3 & 36.9
& 32.4 \\

RGS
& \textbf{54.6} & \textbf{49.4} & \textbf{37.5} & \underline{46.3} & \underline{31.0} & \textbf{33.1} & \underline{39.0}
& \underline{19.4}& 23.6
& \underline{5.4}& \underline{32.2}& \underline{38.1}
& \underline{34.1} \\

REPAIR
& 51.9 & 46.8 & 33.9 & 43.5 & 30.4 & 29.3 & \textbf{39.1}
& 16.3 & 24.1
& 4.6 & 30.9 & 37.1
& 32.3 \\

(Ours) \ours{}
& \underline{53.7} & \underline{48.1} & \underline{37.1} & \textbf{47.9} & \textbf{35.8} & \underline{32.9} & \underline{39.0}
& \textbf{19.7}& 24.2
& \textbf{6.3}& \textbf{33.6}& \textbf{40.0}
& \textbf{34.9} \\
\bottomrule
\end{tabular}
}
\caption{
nDCG@10 performance of various methods on BRIGHT under the reasoning reranking setting, using the top-100 ReasonIR results with GPT-4 queries.
The best performance is marked in \textbf{bold}, and the second-best is \underline{underlined} within the reasoning-based reranking setting.
}
\label{tab:appendix_reasonir_ndcg}
\end{table*}

\begin{table*}[t]
\centering
\renewcommand{\arraystretch}{1.2}
\resizebox{\linewidth}{!}{
\begin{tabular}{l|ccccccc|cc|ccc|c}
\toprule
\multicolumn{14}{l}{\textit{\textbf{Corpus graph embeddings:} Qwen3-Embedding-4B $\rightarrow$ BGE-Large}} \\
\midrule
\multirow{2}{*}{\textbf{Method}}
& \multicolumn{7}{c|}{\textbf{StackExchange}}
& \multicolumn{2}{c|}{\textbf{Coding}}
& \multicolumn{3}{c|}{\textbf{Theorem-based}}
& \multirow{2}{*}{\textbf{Avg.}} \\
\cmidrule(lr){2-8}
\cmidrule(lr){9-10}
\cmidrule(lr){11-13}
& Bio. & Earth. & Econ. & Psy. & Rob. & Stack. & Sus.
& Leet. & Pony
& AoPS & TheoQ. & TheoT. & \\
\midrule
BM25
& 18.9 & 27.2 & 14.9 & 12.5 & 13.6 & 18.4 & 15.0
& 24.4 & 7.9
& 6.2 & 10.4 & 4.9
& 14.5 \\
\midrule
\multicolumn{14}{l}{\cellcolor{gray!15}\textit{Non-Reasoning-based Reranking}} \\
\midrule
Retrieve-and-Rerank
& 34.0 & \underline{41.6} & 21.7 & 28.7 & 29.7 & 20.2 & 29.9
& \underline{17.5} & 14.5
& 3.6 & 13.7 & 14.8
& 22.5 \\
SlideGAR
& 42.4 & 41.4 & 26.5 & 31.8 & 27.9 & 20.9 & 33.5
& 15.2 & \underline{14.6}
& \textbf{4.4} & 12.1 & 17.6
& 24.0 \\
RGS
& \underline{43.3} & \textbf{42.8} & \underline{27.1} & \underline{36.3} & \underline{30.1} & \underline{21.9} & \underline{34.3}
& \textbf{21.3} & 14.2
& 3.7 & \underline{14.5} & \underline{23.8}
& \underline{26.1} \\
(Ours) \ours{}
& \textbf{44.0} & 41.1 & \textbf{29.4} & \textbf{37.2} & \textbf{32.5} & \textbf{25.1} & \textbf{34.3}
& 15.6 & \textbf{27.0}
& \underline{4.2} & \textbf{16.1} & \textbf{33.7}
& \textbf{28.3} \\
\midrule
\multicolumn{14}{l}{\cellcolor{gray!15}\textit{Reasoning-based Reranking}} \\
\midrule
Retrieve-and-Rerank
& 34.7 & 42.5 & 23.8 & 29.0 & 31.6 & 23.2 & 32.0
& \textbf{23.4} & \underline{24.1}
& \underline{5.7} & 15.7 & 13.4
& 24.9 \\
SlideGAR
& 40.2 & 42.3 & 28.3 & 31.0 & 30.2 & 27.6 & 35.2
& 21.3 & 23.2
& \textbf{5.8} & 14.8 & 16.2
& 26.3 \\
RGS
& \textbf{45.2} & \textbf{44.9} & \underline{32.3} & \underline{38.1} & \underline{32.2} & 26.9 & 36.7
& 20.2 & 21.4
& 4.7 & \underline{16.5} & \underline{27.0}
& \underline{28.8} \\
REPAIR
& 41.8 & \underline{44.3} & 31.1 & 33.9 & 31.4 & \underline{29.1} & \underline{37.0}
& \underline{23.0} & 21.6
& 5.1 & 16.0 & 23.7
& 28.2 \\
(Ours) \ours{}
& \underline{43.1} & 44.0 & \textbf{33.1} & \textbf{38.9} & \textbf{33.0} & \textbf{30.1} & \textbf{37.4}
& 20.3 & \textbf{35.6}
& 5.7 & \textbf{17.4} & \textbf{30.3}
& \textbf{30.7} \\
\bottomrule
\end{tabular}
   }
\caption{
nDCG@10 on BRIGHT when constructing the corpus graph with BGE-Large embeddings.
The best performance is marked in \textbf{bold}, and the second-best is \underline{underlined} within the same reranking prompt setting.
}
\label{tab:bge_results}
\vskip -0.2in   
\end{table*}

\noindent\textbf{Corpus Graph Embedding.}
We also evaluate whether \ours{} is robust to the embedding model used for corpus graph construction by replacing Qwen3-Embedding-4B with BGE-Large~\cite{BGE-LARGE}.
As shown in Table~\ref{tab:bge_results}, \ours{} continues to outperform all baselines on BRIGHT under both prompting settings, indicating that the gains of group-aware adaptive retrieval do not rely on a particular embedding model.

\noindent\textbf{Scaling the Reranker and the Navigator.}
We analyze how the capacity of each component affects performance. To this end, we vary the size of the reranker and the navigator across three scales within each of the Qwen3.5~\cite{qwen3.5} and Gemma4~\cite{gemmateam2026gemma4} model 
families, and measure nDCG@10 and latency for all combinations. Since these models are not trained for pointwise relevance judgment, we obtain navigator scores using the same prompt as Qwen3-Reranker. As in Section~\ref{sec:efficiency}, we measure latency on 30 queries sampled from each of the 12 BRIGHT subsets.
As shown in Figure~\ref{fig:scaling}, scaling the reranker leads to substantially larger performance gains than scaling the navigator. This reflects the division of roles between the two components: while the navigator determines which directions are explored, promoting the exposed documents into the top ranks remains the reranker's responsibility. The latency comparison points the same way. Pairing a small navigator with a larger reranker is both faster and more accurate than the reverse. Direction assessment at the group level is thus a less demanding task than fine-grained relevance assessment at the document level, supporting our design choice of a lightweight navigator.
\begin{figure*}[t]
\centering
\includegraphics[width=\textwidth]{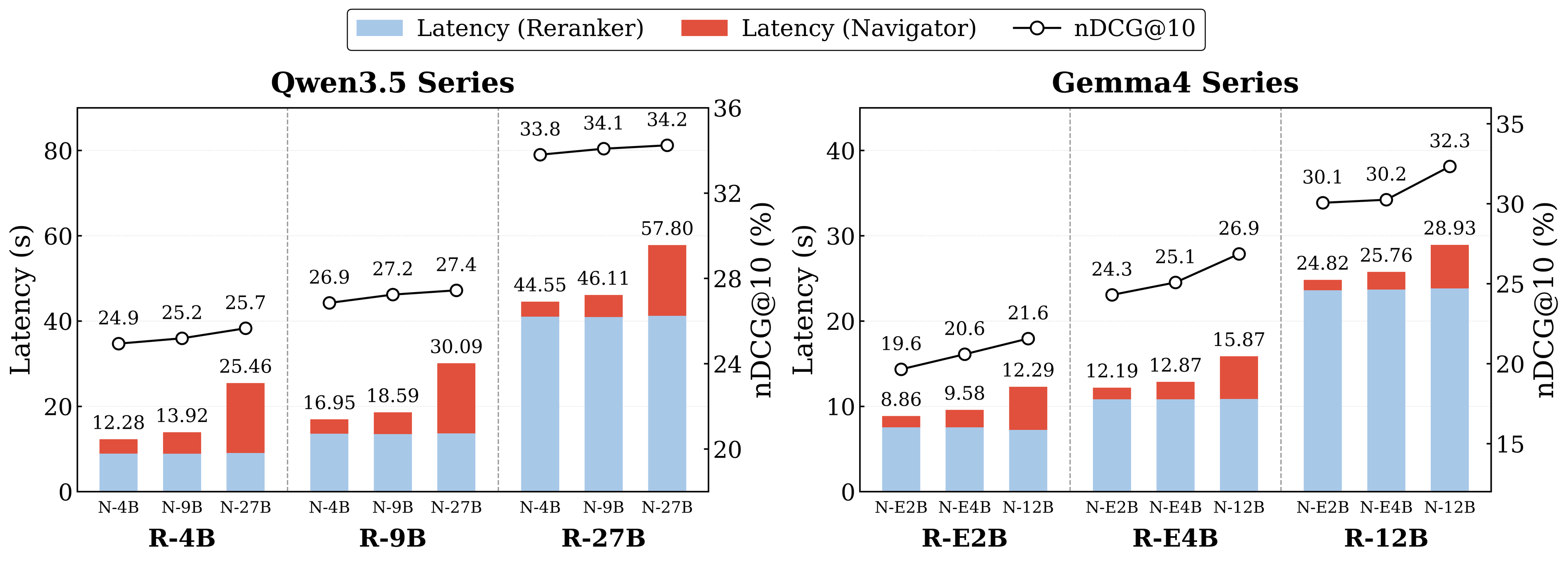}
\caption{%
nDCG@10 and latency across reranker (R) and navigator (N) sizes on BRIGHT. Bars decompose latency into the reranker and the navigator components.
}
\label{fig:scaling}
\vspace{-0.1in}
\end{figure*}

\noindent\textbf{Reranker Backbones.}
To verify that \ours{} is not tied to a specific reranker, we replace the reranker with GPT-4o-mini and Gemma4-26B-A4B-it, and evaluate all methods under the same backbone for a controlled comparison.
As shown in Table~\ref{tab:tab_GPT_GEMMA}, \ours{} achieves the strongest performance in most domain categories across both backbones and under both prompting settings, indicating that the benefit of group-level expansion is not tied to a particular reranker backbone.
\begin{table}[t]
\centering
\small
\setlength{\tabcolsep}{4pt}
\renewcommand{\arraystretch}{1.15}
\begin{tabular}{lcccccc}
\toprule
\multirow{2}{*}{\textbf{Method}}
& \multicolumn{3}{c}{\textbf{GPT}}
& \multicolumn{3}{c}{\textbf{Gemma4}} \\
\cmidrule(lr){2-4}
\cmidrule(lr){5-7}
& SE & CD & TH & SE & CD & TH \\
\midrule
BM25
& 17.2 & 16.1 & 7.1
& 17.2 & 16.1 & 7.1 \\
\midrule
\multicolumn{7}{l}{\cellcolor{gray!15}\textit{Non-Reasoning Reranking}} \\
\midrule
Retrieve-and-Rerank
& 25.4 & 21.4 & 11.9
& 32.9 & \underline{29.1} & 14.4 \\
SlideGAR
& \underline{27.0} & 19.3 & 14.2
& 37.2 & 26.0 & 15.3 \\
RGS
& \underline{27.0} & \underline{21.6} & \underline{15.7}
& \underline{37.3} & 26.1 & \underline{20.6} \\
(Ours) \ours{}
& \textbf{30.6} & \textbf{24.8} & \textbf{19.3}
& \textbf{39.3} & \textbf{30.6} & \textbf{23.1} \\
\midrule
\multicolumn{7}{l}{\cellcolor{gray!15}\textit{Reasoning Reranking}} \\
\midrule
Retrieve-and-Rerank
& 27.4 & \underline{26.0} & 12.4
& 33.5 & \textbf{34.4} & 13.7 \\
SlideGAR
& 29.0 & 23.0 & 14.6
& 38.6 & 29.6 & 16.1 \\
RGS
& \underline{29.2} & 23.9 & \underline{18.8}
& \underline{40.5} & 26.2 & \underline{19.4} \\
REPAIR
& 28.7 & 25.3 & 15.8
& 38.8 & 29.3 & 17.6 \\
(Ours) \ours{}
& \textbf{32.2} & \textbf{27.6} & \textbf{20.5}
& \textbf{42.3} & \underline{30.5} & \textbf{23.9} \\
\bottomrule
\end{tabular}
\caption{
nDCG@10 on BRIGHT grouped by domain categories.
SE, CD, and TH denote StackExchange, Coding, and Theorem-based, respectively.
The best and second-best performance within each reranking setting are marked in \textbf{bold} and \underline{underlined}, respectively.
}
\label{tab:tab_GPT_GEMMA}
\vskip -0.1in   
\end{table}

\subsection{Comparison with Hierarchical Retrieval}

LATTICE~\cite{LATTICE} performs retrieval by organizing the entire corpus into a semantic tree and having an LLM traverse the hierarchy top-down, from high-level abstractions to leaf documents.
In contrast, \ours{} starts from an initial document retrieval state and expands it iteratively: the group-level navigator evaluates which surrounding semantic directions to explore, while the document-level reranker assesses the concrete documents reached along those directions.
Thus, LATTICE uses coarse-to-fine hierarchical traversal as the retrieval process, whereas \ours{} couples coarse-grained directional guidance with fine-grained document feedback for adaptive expansion.

For LATTICE, we use the official pre-built semantic trees released by the authors.
Since both methods construct their corpus structures offline before retrieval, we compare only the LLM usage incurred during iterative retrieval.
We estimate token cost using the input and output token counts of each LLM call during retrieval and the corresponding per-token prices reported by OpenRouter\footnote{\url{https://openrouter.ai/}}.
For a controlled comparison, we stop LATTICE after it has examined 100 documents, matching the document observation budget used in \ours{}.
Table~\ref{tab:lattice_comparison} reports retrieval effectiveness and LLM usage and cost incurred during iterative retrieval on BRIGHT.
LATTICE achieves strong retrieval effectiveness, but its LLM-driven traversal incurs substantially higher generation cost.
With BM25 as the first-stage retriever, \ours{} achieves slightly lower ranking performance while requiring substantially lower iterative-retrieval cost.
When using an enhanced setup with GPT-4 queries and ReasonIR, \ours{} achieves stronger retrieval effectiveness while maintaining lower iterative-retrieval cost than LATTICE.
These results suggest that group-level adaptive expansion with document-level feedback can provide an effective and efficient alternative to top-down hierarchical traversal.
\begin{table}[t]
\centering
\renewcommand{\arraystretch}{1.2}
\resizebox{\columnwidth}{!}{
\begin{tabular}{lccccc}
\toprule
\textbf{Method} & \textbf{nDCG@10} & \textbf{R@100} & \textbf{\# Proc.} & \textbf{\# Gen.} & \textbf{Cost (\$)} \\
\midrule
LATTICE         & 31.9 & 49.2 & 85,419 & 13,859 & 0.0229 \\
\midrule
\ours{} (BM25)       & 31.2 & 52.9 & 125,031 & 4,683 & 0.0109 \\
\quad\quad \textit{Reranker}   & --   & --   & 48,938 & 4,523 & 0.0094 \\
\quad\quad \textit{Navigator}  & --   & --   & 76,093 & 160   & 0.0015 \\
\midrule
\ours{} (ReasonIR)   & {34.9} & {62.9} & 121,994 & 4,684 & {0.0106} \\
\quad\quad \textit{Reranker}   & --   & --   & 45,880 & 4,524 & 0.0091 \\
\quad\quad \textit{Navigator}  & --   & --   & 76,114 & 160   & 0.0015 \\
\bottomrule
\end{tabular}
}
\caption{
Overall performance and online iterative retrieval LLM cost comparison of \ours{} with LATTICE on BRIGHT.
}
\label{fig:scaling}
\label{tab:lattice_comparison}
\end{table}

\section{Analysis of Group-Level Expansion}

\subsection{Structural Analysis of the Group Unit}\label{sec:DOC_GROUP_ANALYSIS}

Section~\ref{sec:DOC_GROUP} showed that group-level expansion surpasses document-level expansion as iterations proceed. In this section, we analyze how the two expansion strategies differ in their ability to reach gold documents at varying distances, and provide a structural explanation for this gap by comparing the distance to gold documents on the document graph and the group graph.

\begin{figure}[t]
\includegraphics[width=1.0\linewidth]{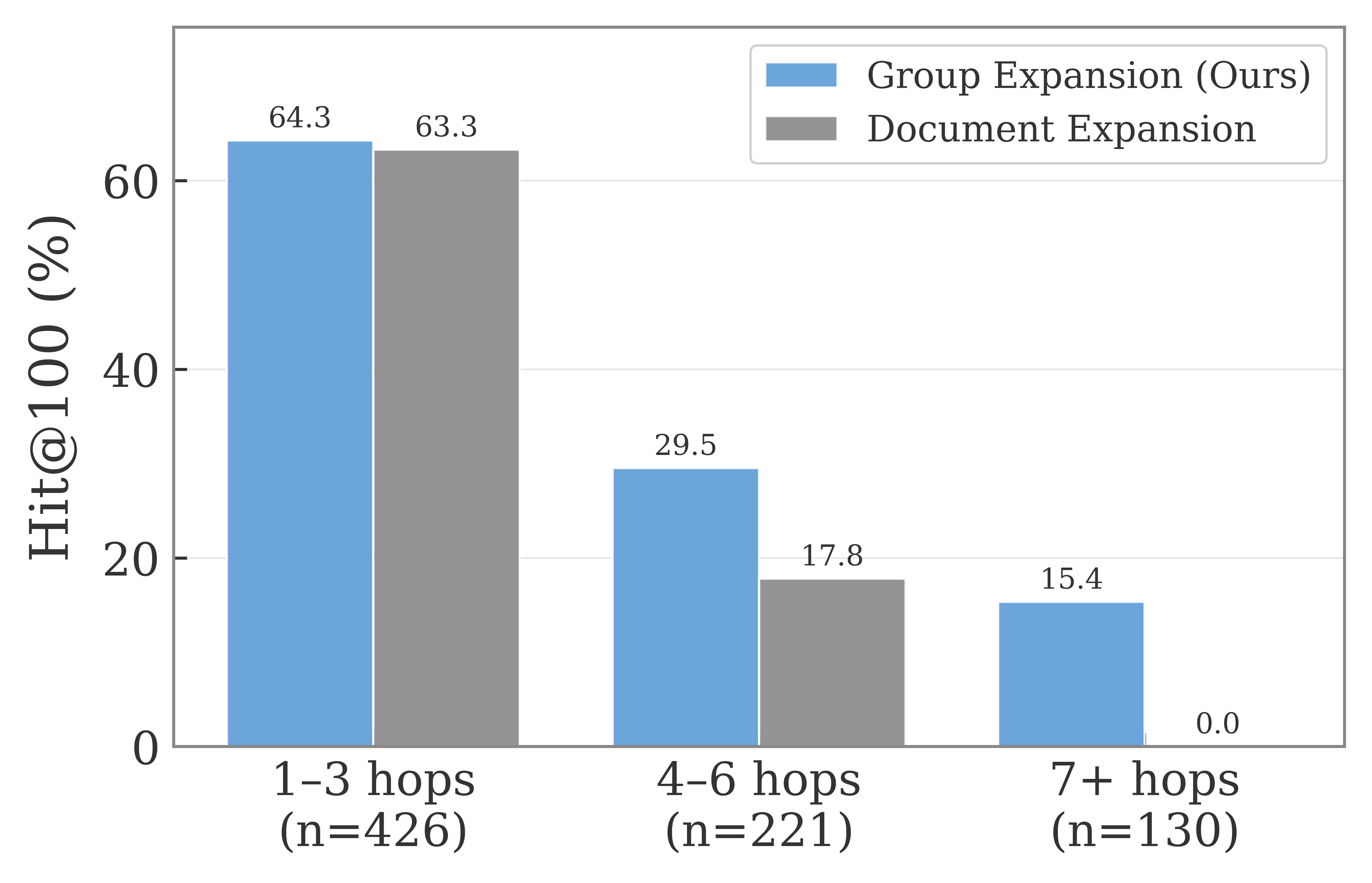}

\caption{%
Hit@100 comparison between group-level and document-level expansion by graph-hop distance to the nearest gold document on BRIGHT. The gap between the two widens as gold documents become farther from the initial retrieval results.
}
\label{fig:doc_group_hit}
\vskip -0.1in
\end{figure}

\noindent\textbf{Reachability by distance.}
We measure how the reachability of the two expansion strategies varies with the distance to the gold document. Following the setting in Section~\ref{sec:gold_location}, we group queries by the minimum document-hop distance from the initial retrieval set to a gold document and report Hit@100. The comparison target is the same document-level expansion variant as in Section~\ref{sec:DOC_GROUP}, keeping all components identical except for the expansion unit. As shown in Figure~\ref{fig:doc_group_hit}, the two strategies show little difference at short distances, but the gap widens as the distance increases. Document-level expansion fails to reach any gold document at 7 or more hops, whereas group-level expansion maintains meaningful reachability even at long distances. This indicates that the performance gap between the two strategies stems from their ability to recover gold documents far from the initial set.

\begin{figure}[t]
\includegraphics[width=1.0\linewidth]{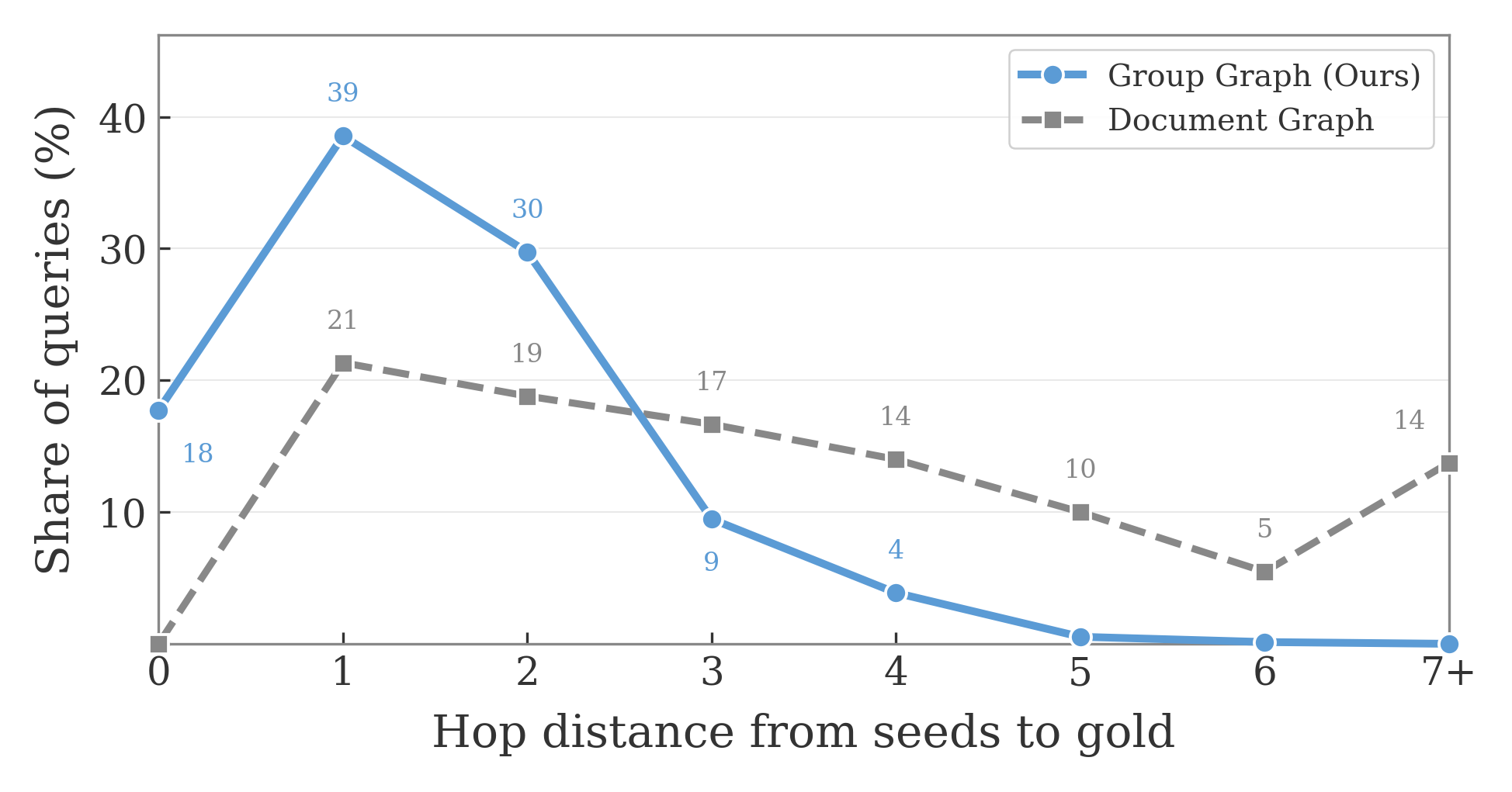}
\caption{%
Distribution of the graph-hop distance to gold documents on the document graph and the group graph on BRIGHT. Distances are measured to the nearest gold document missed by first-stage retrieval, which must be reached through expansion.
}\label{fig:hop_hist_bm25}
\vskip -0.1in
\end{figure}

\noindent\textbf{Distance to gold documents.}
To analyze the structural cause of this gap, we focus on gold documents that are missed by first-stage retrieval and must therefore be reached through expansion. For each query, we measure the minimum hop distance from the initial retrieval set to such a gold document on the document graph and the group graph, respectively. We connect two groups with a directed edge when a document in one group points to a document in the other. As shown in Figure~\ref{fig:hop_hist_bm25}, gold documents are spread across a wide range of distances on the document graph, whereas most of them are concentrated within 3 hops on the group graph. As a result, group-level expansion can skip multiple document hops with a single group selection, reaching distant gold documents within a limited iteration budget.

\subsection{Group Construction Strategies}
\begin{table}[t]
\centering
\renewcommand{\arraystretch}{1.2}
\resizebox{\columnwidth}{!}{
\begin{tabular}{l|cc}
\toprule
\multirow{2}{*}{\textbf{Group Construction}} & \multicolumn{2}{c}{{BRIGHT Avg.}} \\
\cmidrule(lr){2-3}
& nDCG@10 & R@100 \\
\midrule
Random Group              & 23.1 & 31.2 \\
K-means Group             & 26.7 & 50.2 \\
\midrule
Community Detection (Ours) & \textbf{28.3} & \textbf{53.1} \\
\bottomrule
\end{tabular}
}
\caption{
Overall performance of different group construction methods in \ours{}.
}
\label{tab:group_ablation}
\vskip -0.1in
\end{table}

We further compare strategies for constructing document groups. To control for the number of groups, we set the number of K-means clusters equal to the number of groups produced by our community detection construction, and randomly partition documents into the same number of groups for random grouping. As shown in Table~\ref{tab:group_ablation}, community detection performs best, suggesting that graph connectivity provides useful structure for defining expansion directions. Moreover, random grouping falls far behind both methods, showing that coarsening the graph alone brings no benefit. These results confirm that the effectiveness of the group unit comes from the meaningful signals that semantically coherent groups provide.

\subsection{Failure Case Analysis}
\label{sec:failure_analysis}

To understand why GAREN fails to expand gold documents, we analyze the retrieval process with respect to the \textit{gold group}, the group containing the gold documents for a query. We first examine whether the gold group was scored by the navigator during retrieval. With BM25 first-stage retrieval, the gold group was never scored in 28.5\% of the pairs. Stronger first-stage retrieval with ReasonIR and GPT-4 queries reduces this rate to 20.4\% but does not eliminate such cases. Because these failures accumulate over sequential expansion decisions, they are difficult to attribute to specific causes. We therefore focus on the cases where the gold group was scored but not selected, qualitatively analyzing 10 sampled cases from each of the 12 BRIGHT subsets (120 in total) and reporting the proportion of each failure type. Representative examples are presented in Table~\ref{tab:case-study}.
\begin{enumerate}[leftmargin=*,topsep=2pt,itemsep=2pt]
\item \textbf{Failures in group construction (53.3\%).}
The gold document is not properly represented during group construction or summarization, in two ways. First, even when the document is assigned to a semantically appropriate group, the summary is generated around the content shared across the group and omits aspects specific to individual documents. Second, the corpus graph connects documents through surface-level term overlap rather than topical similarity, assigning the document to a group unrelated to its topic.

\item \textbf{Failures in the navigator's judgment (46.7\%).}
The group assignment is appropriate and the summary contains relevant information, yet the navigator assigns a low score. This is pronounced when the relevance between the query and the summary requires multi-step reasoning.
\end{enumerate}
\noindent\textbf{Improvement directions.}
The two categories suggest different directions for improvement. Failures in group construction call for better grouping methods, such as constructing groups from LLM-extracted topics of each document, or soft clustering that represents the multiple aspects of a document across different groups. For failures in the navigator's judgment, Figure~\ref{fig:scaling} shows that a larger navigator improves group selection, but the latency cost outweighs the gains, making scale an inefficient solution. Training the navigator to infer the topics required by a query could be a mitigation. We leave these directions as future work.

\section{Hyperparameter Sensitivity}
In this section, we analyze the sensitivity of \ours{} to the switching threshold and the navigator budget.

\noindent\textbf{Explore-then-Exploit Strategy.}
\begin{figure}[t]
\includegraphics[width=1.0\linewidth]{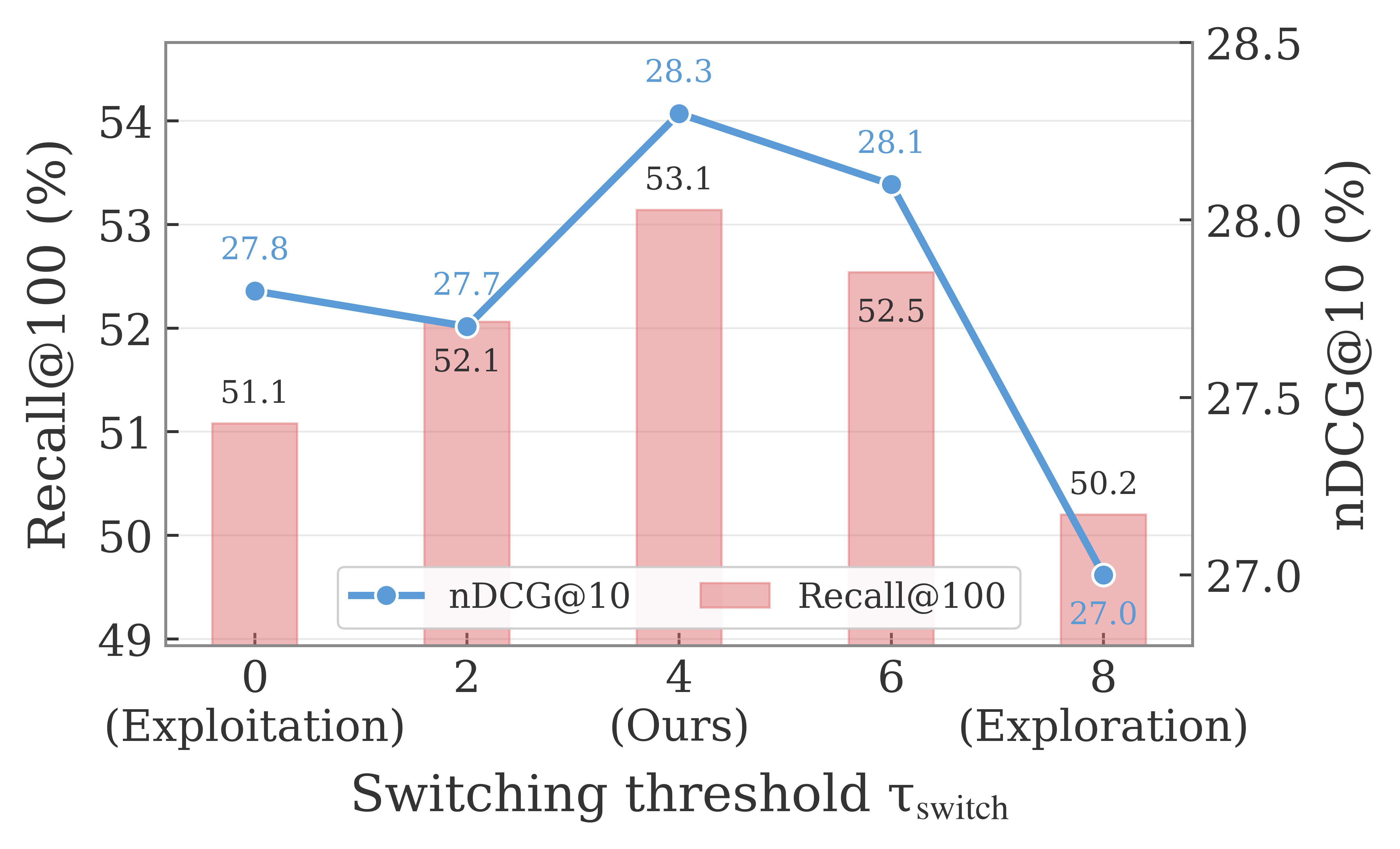}
\caption{
Performance under different switching thresholds $\tau_{\mathrm{switch}}$ on BRIGHT.
}\label{fig:transition}
\vskip -0.1in
\end{figure}

Figure~\ref{fig:transition} shows the effect of varying $\tau_{\mathrm{switch}}$, the iteration at which \ours{} switches from exploration to exploitation.
Both extremes underperform: $\tau_{\mathrm{switch}}=0$ (exploitation) commits too early to a narrow set of groups, while $\tau_{\mathrm{switch}}=8$ (exploration) spreads expansion across too many groups without sufficient focus.
Performance peaks at $\tau_{\mathrm{switch}}=4$, which we use as our default, confirming that the explore-then-exploit strategy benefits from a balanced transition.

\begin{figure}[t]
\includegraphics[width=1.0\linewidth]{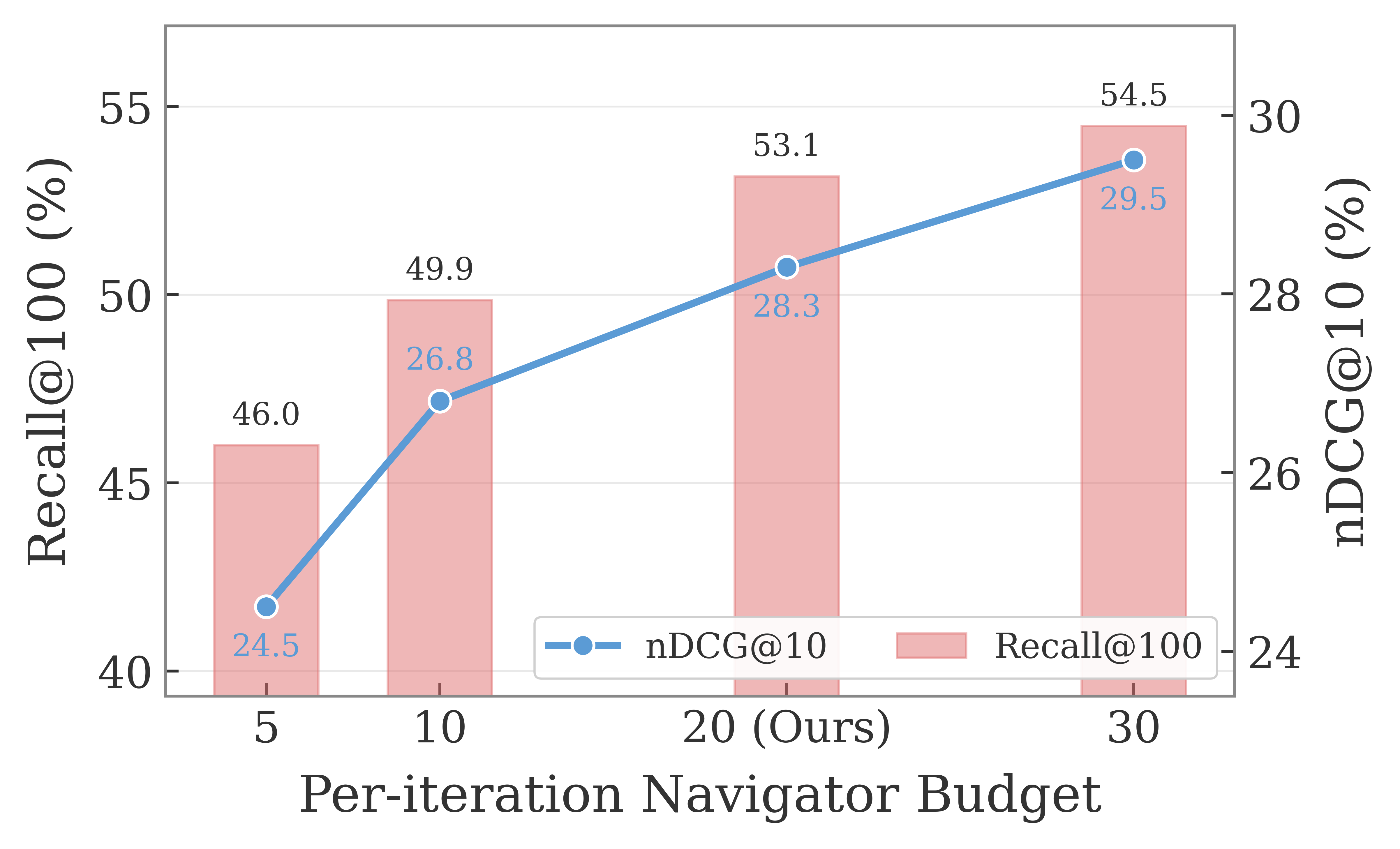}
\caption{
Performance under different per-iteration navigator budgets $K_g$ on BRIGHT.
}\label{fig:navigator_budget}
\vskip -0.1in
\end{figure}

\noindent\textbf{Navigator Budget.}
Figure~\ref{fig:navigator_budget} shows the effect of varying $K_g$, the maximum number of groups scored by the navigator per iteration.
Performance improves with larger $K_g$, since scoring more groups expands the pool of expansion candidates available to \ours{}.
We adopt $K_g=20$ as a balance between accuracy and computational cost; while further increases yield additional gains, they incur proportionally higher navigator cost.

\section{Details of the Online Retrieval Procedures}

This section provides the detailed definitions used in candidate group selection.
At iteration $t$, let $W_t$ be the current document window, let $N_k(d)$ denote the $k$-NN neighbors of document $d$ on the corpus graph, and let $\Omega(d) \in \mathcal{P}$ denote the unique group containing document $d$.

\paragraph{Navigate.}
\label{app:voting_caching}

This section details how the navigator scores the candidate groups $\mathcal{G}^{\mathrm{cand}}_t$ (Section~\ref{sec:retrieval}) under a per-iteration budget $K_g$ using three rules.

\begin{enumerate}[leftmargin=*,topsep=2pt,itemsep=2pt]
\item \textbf{Reuse.} A navigator score depends only on the query and the group, not on the iteration, so we cache it in $\Pi_q$. A candidate group already scored in an earlier iteration reuses its cached score $\pi_q(C_m)$ without recomputation.

\item \textbf{Score.}
At each iteration, the navigator newly scores at most $K_g$ previously unscored candidate groups.
For each candidate group $C_m$, a pointwise relevance model evaluates the query $q$ against the group summary $S_m$.
Although the model is prompted to answer with \texttt{yes} or \texttt{no}, we define $\Phi(q,S_m)$ as the softmax-normalized probability assigned to \texttt{yes} over the two response logits.
The group score is computed as:
\begin{equation}
    \pi_q(C_m) = \Phi(q, S_m).
\end{equation}
Candidate groups are ranked in descending order of $\pi_q(C_m)$, and each computed score is cached for reuse in subsequent iterations.
The navigator prompt is provided in Table~\ref{tab:prompt_all}.

\item \textbf{Budget allocation.} When more than $K_g$ unscored candidates are available, we score the $K_g$ groups with the highest \emph{vote}, the number of window documents whose neighborhood intersects the group:
\begin{equation}
\texttt{vote}_t(C_m) = \sum_{d \in \mathcal{W}_t} \mathbb{1}\!\left[\,N_k(d) \cap C_m \neq \emptyset\,\right].
\end{equation}

When fewer than $K_g$ are available, we fill the remaining budget by scanning the first-stage pool in rank order and adding the groups of the top-ranked uncovered documents. Groups that remain unscored under the budget are excluded from expansion in the current iteration, and may be scored in a later one.
\end{enumerate}

\noindent
Only groups with a cached or newly computed score are eligible for expansion in the current iteration, forming $\mathcal{G}^{\mathrm{score}}_t$.

\noindent\textbf{Window Expansion.}
Given the scored candidate groups $\mathcal{G}^{\mathrm{score}}_t$, we construct the expansion set $\mathcal{X}_t$ according to the explore-then-exploit strategy.
Within each visited group, we select unobserved documents in descending order of their rank in the first-stage retrieval result until the expansion budget is filled.
The resulting expansion set is combined with the retained reranked documents to construct the next retrieval window.

\section{Group Statistics}
\begin{table}[t]
\centering
\renewcommand{\arraystretch}{1.2}
\resizebox{\columnwidth}{!}{
\begin{tabular}{l|rr|rrr|r}
\toprule
\multirow{2}{*}{\textbf{Dataset}}
& \multirow{2}{*}{\textbf{\# Docs}}
& \multirow{2}{*}{\textbf{\# Groups}}
& \multicolumn{3}{c|}{\textbf{Group Size}}
& \multirow{2}{*}{\textbf{Time (mins)}} \\
\cmidrule(lr){4-6}
 & & & Min & Max & Mean & \\
\midrule
\multicolumn{7}{l}{\textit{StackExchange}} \\
\midrule
Biology              &  57,359 &  5,183 & 2 &  58 & 11.1 &  81 \\
Earth Science        & 121,249 & 11,485 & 2 &  22 & 10.6 & 196 \\
Economics            &  50,220 &  4,561 & 2 & 194 & 11.0 &  64 \\
Psychology           &  52,835 &  4,789 & 2 & 240 & 11.0 &  65 \\
Robotics             &  61,961 &  4,861 & 2 & 671 & 12.7 &  70 \\
Stack Overflow       & 107,081 &  8,136 & 2 & 672 & 13.2 & 149 \\
Sustainable Living   &  60,792 &  5,540 & 2 & 208 & 11.0 &  76 \\
\midrule
\multicolumn{7}{l}{\textit{Coding}} \\
\midrule
LeetCode             & 413,932 & 40,861 & 2 &  19 & 10.1 & 611 \\
Pony                 &   7,894 &    621 & 2 &  94 & 12.7 &  11 \\
\midrule
\multicolumn{7}{l}{\textit{Theorem-based}} \\
\midrule
AoPS                 & 188,002 & 16,558 & 2 &  22 & 11.4 & 364 \\
TheoremQA-Q          & 188,002 & 16,558 & 2 &  22 & 11.4 & 364 \\
TheoremQA-T          &  23,839 &  2,196 & 2 &  19 & 10.9 &  40 \\
\bottomrule
\end{tabular}
}
\caption{
Group construction statistics for \ours{} on BRIGHT.
Time is reported in minutes.}
\label{tab:group_stats}
\end{table}

Group construction is performed offline once for each corpus before retrieval.
Table~\ref{tab:group_stats} reports the number and size distribution of the resulting groups, along with the time required for graph partitioning and group summary generation on each BRIGHT dataset.
Graph partitioning completes within one minute for each dataset, while the reported summary generation times are measured using two NVIDIA A100 GPUs.

\noindent\textbf{Analysis of Large Groups.}
Although we set the re-partitioning target threshold to $\bar{s}=20$, a few groups exceed this limit.
This occurs because some documents in the corpus are effectively identical (or near-identical), making them inseparable during graph partitioning. As a result, these groups remain larger than the target threshold.

\noindent\textbf{Scalability.}
We also confirmed that GAREN operates on corpora of up to 4.6M documents (DBpedia in BEIR) without modification, where the offline group construction cost grows roughly linearly with corpus size. The online cost also remains bounded regardless of corpus size, since the navigator scores at most $K_g$ groups per iteration.

\section{Prompts}
\label{app:prompts}

\noindent\textbf{Navigator and Reranker Prompts.}
Table~\ref{tab:prompt_all} presents the prompts used for the group-level navigator and the 
document-level reranker in \ours{}. We adopt two reranking prompts, non-reasoning and reasoning, for two reasons. First, REPAIR guides expansion using intermediate reasoning steps and is thus applicable only under the reasoning prompt setting, which is required for a fair comparison with it. Second, the two prompts differ substantially in cost, as the reasoning prompt generates a reasoning trace at every iteration (Figure~\ref{fig:latency_accuracy_curve}). Reporting both settings makes this effectiveness-efficiency trade-off explicit.

\noindent\textbf{Group Summary Generation.}
Table~\ref{tab:prompt_group_summary} presents the prompt used for summary generation.
Although the prompt generates multiple descriptive fields for a group, we use only the generated \textit{concept} and \textit{summary} as its representation for subsequent group-level navigation.
Table~\ref{tab:summary-example-size5} provides an illustrative example of a generated group representation.

\begin{table*}[t]
\centering
\renewcommand{\arraystretch}{1.06}
\setlength{\tabcolsep}{6pt}

\begin{tabularx}{\textwidth}{
    @{}
    >{\raggedright\arraybackslash}p{0.19\textwidth}
    @{\hspace{7pt}\vrule width 0.4pt\hspace{7pt}}
    Z
    @{}
}

\toprule

\multicolumn{2}{@{}l@{}}{\textit{\textbf{Reranker (Reasoning)}}} \\
\midrule

\textbf{System prompt}
&
\PromptText{
You are an AI assistant that analyzes complex questions and identifies which documents best support answering them.
\newline\newline
Given a user's query and a set of documents, your task is to:
\newline
1. Generate a reasoning trace, thinking step by step about what knowledge or types of information are necessary to answer the query. These should be abstract but specific enough to guide document selection.
\newline
2. Select and rank at least 10 documents that best support the reasoning steps. Consider how each document contributes to the reasoning process. Order them from most to least useful using > between document IDs (e.g., [3] > [7]).
\newline\newline
Use the following format:
\newline
[Reasoning Trace]
\newline
Step 1: <First reasoning step>
\newline
Step 2: <Second reasoning step>
\newline
...
\newline
Step N: <Final reasoning step>
\newline\newline
[Document Ranking]
\newline
[9] > [5] > [6] > ... > [12]
\newline\newline
Only produce the output in the format shown above.
}
\\
\cmidrule{1-2}

\textbf{User prompt}
&
\PromptText{
[Query]
\newline
\{query\}
\newline\newline
[Documents]
\newline
\{docs\}
}
\\

\midrule
\multicolumn{2}{@{}l@{}}{\textit{\textbf{Reranker (Non-reasoning)}}} \\
\midrule

\textbf{System prompt}
&
\PromptText{
You are RankLLM, an intelligent assistant that can rank documents based on their relevancy to the query.
}
\\
\cmidrule{1-2}

\textbf{User prompt}
&
\PromptText{
I will provide you with passages, each indicated by number identifier [].
Rank the passages based on their relevance to the search query.
\newline
\{docs\}
\newline\newline
Search Query: \{query\}.
\newline
Rank the \{n\_docs\} passages above based on their relevance to the search query.
\newline\newline
The passages should be listed in descending order using identifiers.
The most relevant passages should be listed first.
The output format should be <answer> [] > [] </answer>,
e.g., <answer> [1] > [2] </answer>.
}
\\

\midrule
\multicolumn{2}{@{}l@{}}{\textit{\textbf{Navigator}}} \\
\midrule

\textbf{System prompt}
&
\PromptText{
Judge whether the Document meets the requirements based on the Query and the Instruct provided.
Note that the answer can only be ``yes'' or ``no''.
}
\\
\cmidrule{1-2}

\textbf{User prompt}
&
\PromptText{
<Instruct>: Given a query, retrieve relevant passages that answer the query.
\newline\newline
<Query>: \{query\}
\newline\newline
<Document>: \{doc\}
}
\\

\bottomrule
\end{tabularx}
\caption{Prompt templates used for the reranker and navigator.}
\label{tab:prompt_all}
\end{table*}

\begin{table*}[t]
\centering
\renewcommand{\arraystretch}{1.06}
\setlength{\tabcolsep}{6pt}

\begin{tabularx}{\textwidth}{
    @{}
    >{\raggedright\arraybackslash}p{0.19\textwidth}
    @{\hspace{7pt}\vrule width 0.4pt\hspace{7pt}}
    Z
    @{}
}

\toprule

\multicolumn{2}{@{}l@{}}{\textit{\textbf{Summarizer}}} \\
\midrule

\textbf{System prompt}
&
\PromptText{
You are an expert AI analyst and summarizer.
Output valid JSON only --- no markdown, no prose outside the JSON.
}
\\
\cmidrule{1-2}

\textbf{User prompt}
&
\PromptText{
You are an expert AI analyst and summarizer. Your mission is to create a highly
informative and ``discriminative signpost'' for a navigating search agent.
This signpost (a summary) must guide the agent to the correct cluster of nodes
to answer a user's query.
\newline\newline
You will follow a strict, step-by-step cognitive process. You must analyze the
children nodes in a target parent node (the ``Positive Set'').
\newline\newline
\#\# INPUTS
\newline\newline
\#\#\# POSITIVE SET: Information about the target parent node to be summarized
\newline
The following \{n\} documents belong to the same community detected from a corpus graph.
\newline\newline
\{docs\}
\newline\newline
\#\# YOUR TASK \& OUTPUT FORMAT
\newline\newline
Your entire output must be a single, valid JSON object. Inside this JSON you
will follow the 3-step thinking process outlined below, populating each field
as instructed.
\newline\newline
\#\#\# JSON Structure and Instructions:
\newline\newline
\{
\newline
\hspace*{1em}"detailed\_fingerprints": [
\newline
\hspace*{2em}// For EACH document in the POSITIVE SET (target parent node),
\newline
\hspace*{2em}// extract a structured object of its key, queryable facts.
\newline
\hspace*{2em}\{
\newline
\hspace*{3em}"one\_line\_summary": "...", \quad // very information-dense, concise one-line summary
\newline
\hspace*{3em}"key\_entities": ["..."], \quad // a few key entities central to this document
\newline
\hspace*{3em}"genre\_or\_category": ["..."], \quad // a few key genres / categories
\newline
\hspace*{3em}"name": "..." \quad // name the document
\newline
\hspace*{2em}\}
\newline
\hspace*{1em}],
\newline
\hspace*{1em}"common\_theme": "...",
\newline
\hspace*{1em}// Reason deeply about the common themes between the documents
\newline
\hspace*{1em}// in the POSITIVE SET. Focus on the majority theme; do not force
\newline
\hspace*{1em}// outlier items into it.
\newline
\hspace*{1em}"concept": "...",
\newline
\hspace*{1em}// A 1--5 word label naming this community
\newline
\hspace*{1em}// (e.g., "Blue-eyed cats", "Structural coloration",
\newline
\hspace*{1em}// "HF radio communication").
\newline
\hspace*{1em}"summary": "..."
\newline
\hspace*{1em}// Based on step 1 and step 2, write a very information-dense
\newline
\hspace*{1em}// description of this community. Make sure to include all key
\newline
\hspace*{1em}// entities from the majority theme. This summary must be
\newline
\hspace*{1em}// discriminative enough for a search agent to decide whether
\newline
\hspace*{1em}// to explore this community. Aim for 4--8 sentences covering
\newline
\hspace*{1em}// the core topic plus all notable entities.
\newline
\}
\newline\newline
\#\# IMPORTANT GUIDELINES
\newline
- The summary MUST include the specific entities extracted in step 1
(key\_entities). Avoid vague labels like "Various topics" or "Miscellaneous".
\newline
- Focus on the majority theme. If some documents do not align with the main
theme, down-weight them rather than forcing them into the summary.
\newline
- The "concept" is a short label; the "summary" is the detailed description.
\newline
- Output JSON only --- no markdown code fences, no explanation text.
\newline\newline
Your Response (JSON only):
}
\\

\bottomrule
\end{tabularx}
\caption{Prompt template used for the summarizer module.}
\label{tab:prompt_group_summary}
\end{table*}

\FloatBarrier

\begin{table*}[t]
\centering
\small
\setlength{\tabcolsep}{5pt}
\renewcommand{\arraystretch}{1.1}

\begin{tabular}{@{}p{0.20\linewidth}p{0.76\linewidth}@{}}
\toprule
\textbf{Document} \\
\midrule

$d_1$ \quad \texttt{Hand\_washing\_0\_3}
&
The most commonly missed areas are the thumb, wrist, areas between the
fingers, and under fingernails. Artificial nails and chipped nail polish may
harbor microorganisms. There are five critical times during the day where
washing hands with soap is important to reduce fecal--oral transmission of
disease. \emph{\ldots}
\\
\addlinespace[4pt]

$d_2$ \quad \texttt{Hand\_washing\_0\_1}
&
The World Health Organization recommends washing hands for
\textbf{at least 20 seconds}. These include five critical times where washing
hands with soap is important to reduce fecal--oral transmission.
\emph{\ldots}
\\
\addlinespace[4pt]

$d_3$ \quad \texttt{Hand\_washing\_5\_3}
&
The World Health Organization defines \textbf{``Five Moments''} for hand
hygiene: before patient care, after environmental contact, after exposure to
blood or body fluids, before an aseptic task, and after patient care.
\emph{\ldots}
\\
\addlinespace[4pt]

$d_4$ \quad \texttt{Hand\_washing\_0\_30}
&
Antimicrobial soaps may be desirable before surgery or in environments where
antibiotic-resistant organisms are prevalent. Surgical scrubbing requires a
\textbf{tap that can be operated without touching it with the hands}.
\emph{\ldots}
\\
\addlinespace[4pt]

$d_5$ \quad \texttt{Hand\_washing\_1\_2}
&
Moist hands are more easily recontaminated. Commonly missed areas include the
thumb, wrist, spaces between the fingers, and areas under fingernails.
\emph{\ldots}
\\

\midrule
\multicolumn{2}{@{}l}{\textbf{Community Summary}} \\
\midrule

\emph{Concept}
&
\textbf{WHO Hand Hygiene Protocols}
\\
\addlinespace[3pt]

\emph{Summary}
&
This community consolidates WHO guidance on community and clinical hand
hygiene. It covers the five critical times for community handwashing,
frequently missed anatomical areas, contamination risks from artificial nails
and chipped polish, the clinical ``Five Moments,'' antimicrobial soaps in
surgical and high-risk settings, and touch-free faucets for surgical
scrubbing.
\\

\bottomrule
\end{tabular}

\caption{
Example of group summarization. Five overlapping chunks $d_1$--$d_5$ from the
\emph{Hand washing} Wikipedia article are consolidated into a single
community-level summary.
}
\label{tab:summary-example-size5}
\end{table*}

%
%
%
%

\newcommand{\elide}{\ldots}            

\newcommand{\caseblock}[6]{%
  \begin{tabular}{@{}p{0.15\linewidth}p{0.81\linewidth}@{}}
  \toprule
  \multicolumn{2}{@{}l}{\textbf{#1}}\\
  \midrule
  \emph{Query}          & #2 \\
  \addlinespace[3pt]
  \emph{Query intent}   & #3 \\
  \midrule
  \emph{Gold document}  & #4 \\
  \midrule
  \emph{Group summary}  & #5 \\
  \midrule
  \emph{Why it fails}   & #6 \\
  \bottomrule
  \end{tabular}\\[10pt]
}

\begin{table*}[t]
\centering
\small
\setlength{\tabcolsep}{4pt}

\caseblock
  {[1] Group construction --- missing facet in summary \quad\normalfont\itshape (biology, ``why stars disappear'')}
  {Why do stars disappear when I \textbf{look at them}? \elide{}}
  {needs that foveal cones perform poorly in \textbf{dim light}}
  {Cones \elide{} respond differently to light of different wavelengths, and the combination of their responses is responsible for colour vision. Cones function best in relatively bright light \elide{} as opposed to rod cells, which work better in \textbf{dim light}. \elide{} \emph{(two facets: (i) colour vision, (ii) dim-light performance)}}
  {\elide{} three types of cone photoreceptors---S/M/L-cones---each expressing distinct opsins that determine their \textbf{spectral sensitivity} \elide{}}
  {The group's topic (photoreceptors) matches the query, but the summary describes only facet (i); facet (ii), the one the query needs, is \textbf{missing from the summary}.}

\caseblock
  {[2] Group construction --- topical mismatch in grouping \quad\normalfont\itshape (leetcode, ``max points on a line'')}
  {\elide{} return the maximum number of points that lie on the \textbf{same straight line}.}
  {needs a \textbf{collinearity} counting technique}
  {\texttt{def count\_good\_triplets(nums1, nums2):} \elide{} Return the \textbf{minimum number of lines} needed to represent the line chart. \elide{}}
  {\elide{} \textbf{triplet-counting} problems in arrays, where the goal is to count sets of three elements satisfying combinatorial, geometric, or algorithmic constraints. \elide{}}
  {The document is grouped with triplet problems by a \textbf{surface cue}---its function name---so \textbf{no summary of this group} could surface collinearity.}

\caseblock
  {[3] Navigator judgment  \quad\normalfont\itshape (earth science, ``humidity above 100\%'')}
  {\elide{} what is the limit of humidity and why can it \textbf{exceed 100\%}? \elide{}}
  {needs the \textbf{Kelvin equation}, which explains why vapour stays uncondensed above saturation}
  {The \textbf{Kelvin equation} describes the change in vapour pressure due to a curved liquid--vapor interface, such as the surface of a droplet. \elide{}}
  {The \textbf{Kelvin equation} describes how vapor pressure increases over curved liquid--vapor interfaces due to surface tension \elide{}}
  {Document and summary both name the \textbf{Kelvin equation}, the exact knowledge the query needs, yet the group is still rejected.}

\caption{%
Case study on the failure modes of \ours{}. Failures in group construction appear in two forms: (i) the group topic matches the query but the facet the query needs is not reflected in the summary, and (ii) the gold document is grouped along a surface cue rather than its topic. In failures of navigator judgment (iii), the summary states the required information yet the navigator does not select the group.}
\label{tab:case-study}
\end{table*}

\end{document}